\documentclass[acmtog,authorversion=True]{acmart}

\AtBeginDocument{%
  \providecommand\BibTeX{{%
    \normalfont B\kern-0.5em{\scshape i\kern-0.25em b}\kern-0.8em\TeX}}}

\setcopyright{acmlicensed}
\acmJournal{TOG}
\acmYear{2023} \acmVolume{42} \acmNumber{4} \acmArticle{1} \acmMonth{8} \acmPrice{15.00}\acmDOI{10.1145/3592416}

\usepackage{savesym}
\savesymbol{zifour@default}
\savesymbol{zifour@scaled}
\usepackage{enumerate}
\usepackage{multirow} 
\usepackage[normalem]{ulem}
\usepackage{inconsolata}

\usepackage{xspace}

\newcommand{\rev}[1]{#1}

\newcommand{\revdel}[1]{}

\newcommand{\revf}[1]{#1}

\newcommand{\denselist}{\itemsep 0pt\parsep=0pt\partopsep 0pt\vspace{-\topsep}}

\newcommand{\methodname}{\text{ShapeCoder}\xspace}

\begin{document}

\title[\methodname: Discovering Abstractions for Visual Programs from Unstructured Primitives]{ShapeCoder: Discovering Abstractions for Visual Programs\\ from Unstructured Primitives}

\author{R. Kenny Jones}
\email{russell_jones@brown.edu}
\affiliation{%
    \institution{Brown University}
    \country{USA}
}

\author{Paul Guerrero}
\email{guerrero@adobe.com}
\affiliation{%
    \institution{Adobe Research}
    \country{United Kingdom}
}

\author{Niloy J. Mitra}
\email{n.mitra@cs.ucl.ac.uk}
\affiliation{%
    \institution{University College London and Adobe Research}
    \country{United Kingdom}
}

\author{Daniel Ritchie}
\email{daniel\_ritchie@brown.edu}
\affiliation{%
    \institution{Brown University}
    \country{USA}
}

\begin{abstract}

We introduce \methodname, the first system capable of taking a dataset of shapes, represented with unstructured primitives, and jointly discovering 
(i)~useful \emph{abstraction} functions and 
(ii)~programs that use these abstractions to explain the input shapes.
The discovered abstractions capture common patterns (both structural and parametric) across a dataset, so that programs rewritten with these abstractions are more compact, and suppress spurious degrees of freedom.
\methodname improves upon previous abstraction discovery methods, finding better abstractions, for more complex inputs, under less stringent input assumptions.
This is principally made possible by two methodological advancements: (a)~a shape-to-program recognition network that learns to solve sub-problems and (b)~the use of e-graphs, augmented with a conditional rewrite scheme, to determine when abstractions with complex parametric expressions can be applied, in a tractable manner.
We evaluate \methodname on multiple datasets of 3D shapes, where primitive decompositions are either parsed from manual annotations or produced by an unsupervised cuboid abstraction method.
In all domains, \methodname discovers a library of abstractions that captures high-level relationships, removes extraneous degrees of freedom, and achieves better dataset compression compared with alternative approaches. 
Finally, we investigate how programs rewritten to use discovered abstractions prove useful for downstream tasks.

\end{abstract}



\begin{teaserfigure}
\centering
  \includegraphics[width=\textwidth]{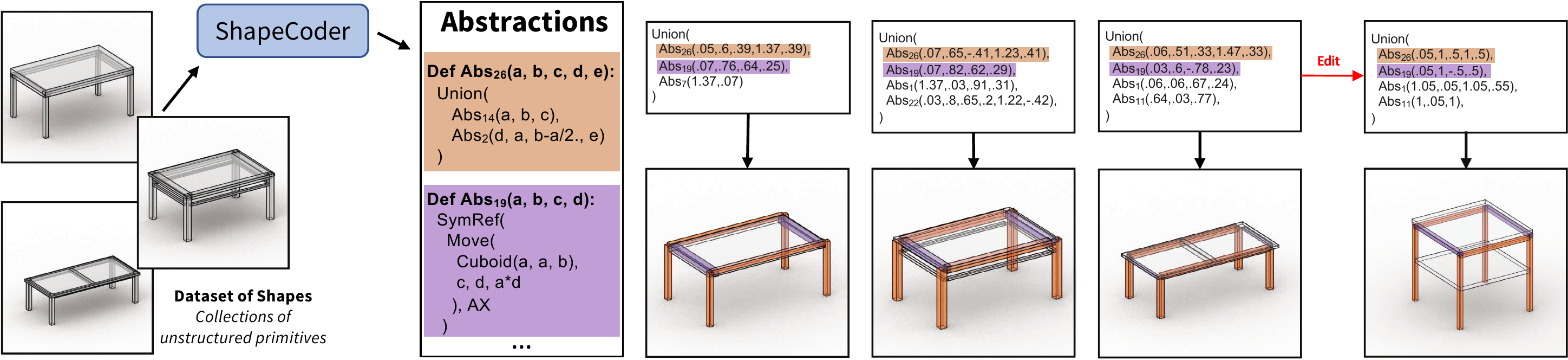}
  \caption{
    \methodname automatically discovers abstraction functions, and infers visual programs that use these abstractions, to compactly explain an input dataset of shapes represented with unstructured primitives.
    For example, the orange abstraction uses only five parameters to encode a distribution of 4-legged table bases with adjoining horizontal support bars.
    }
  \label{fig:teaser}
\end{teaserfigure}

\maketitle

\section{Introduction}

Procedural models are an attractive representation for visual data.
Visual programs, expressions that produce visual outputs when executed, offer many advantages over alternative representations, such as compactness, interpretability, and editability~\cite{NcsgSTAR}. 
There is a wide-range of domains that fall under the purview of visual programs, and procedural workflows are becoming increasingly common in modeling software \cite{mti1040027, houdini}. 
A trait that many visual programming domains share, is that their programs often contain both structural diversity and variables that are constrained by complex parametric relationships.

Typically, visual programs are written in domain-specific languages (DSLs) targeted for specific visual applications.
\revf{
Not all visual programs are equally useful. 
Well-structured programs that capture and constrain properties of the visual data they represent typically benefit downstream applications (e.g. editing, generation, analysis).
On the other hand, badly written programs lose this advantage.
For instance, given an input visual scene composed of a collection of primitives, a visual program that simply unions instantiated primitives together might achieve a perfect reconstruction, but would lose all of the aforementioned benefits of the underlying representation.
The functions a DSL contains influences the types of programs it can represent, and access to a `good' collection of functions is often a prerequisite for finding well-structured programs.
Abstraction functions that extract out common patterns of structural and parametric use for a particular domain, can significantly improve visual program quality, but these types of programs (and their abstractions) are hard to obtain without expert manual design.}

The idea of automatic abstraction discovery has been investigated for general programming domains \cite{DreamCoder}.
Some approaches have also been designed for visual domains \cite{jones2021shapeMOD}, where programs contain complex parametric relationships that complicate this task.
While previous methods have made headway \rev{towards solving this task, none offer a complete solution}.  
Two central limitations holding back the applicability of such methods are that they are either designed without visual programs in mind (so fail to find meaningful parametric relationships) or rely on heavy input assumptions that are hard to meet.

In this paper, we present \methodname, a method that is able to discover useful abstractions for visual data under relaxed assumptions.
\methodname consumes a base DSL and a dataset of shapes represented as collections of primitives without any additional annotations.
\revf{
It discovers a collection of abstraction functions (a library) over the base DSL that is tailored to the input distribution.
It uses the discovered library to find programs with abstractions that explain the shapes from the dataset (Figure \ref{fig:teaser}).
}
Our approach is inspired by, and improves upon, previous abstraction discovery approaches. 

Like DreamCoder~\cite{DreamCoder}, a library learning method for general programming domains, we employ an iterative procedure with interleaved phases (dream, wake, proposal, and integration).
\rev{ These phases are run repeatedly, gradually discovering a library of abstraction functions that minimize a compression-based objective function.}
\rev{The dream phase trains a recognition network, which is used by the wake phase to infer visual programs that explain input shapes.}
Critically, we design our recognition network in a way that allows it to
find partial solutions for difficult input scenes. 
This allows \methodname to still work on input datasets that lack a curriculum of examples (some inputs are easy to solve under the base DSL), which is a limitation of some prior work \cite{DreamCoder}.

\rev{
Using programs from the wake phase, a proposal phase suggests candidate abstractions, which an integration phase reasons over to find improved library versions.
These phases draw inspiration from ShapeMOD~\cite{jones2021shapeMOD}, an abstraction discovery method designed for visual data.
}
ShapeMOD's integration stage relies 
on enumerative search over a limited, curated subset of possible program line-orderings.
\rev{ShapeMOD assumes access to a dataset containing hierarchical part annotations, to keep the number of line-orderings small, but this approach scales poorly.}
To overcome this limitation, we design an integration stage that makes use of e-graphs~\cite{eq_sat_tate, 2021-egg}, a data structure that represents large sets of equivalent programs in a efficient manner through rewrite operations. 
This enables us to search over a huge space of refactored programs,
and allows \methodname to integrate more complex abstractions that  better match the input distribution.
Critical to making e-graph expansion and extraction tractable for our problem setting, 
we implement a novel conditional rewrite scheme 
that lazily evaluates parametric relationships before applying abstraction rewrites, avoiding e-graph size blowup.

We run \methodname over multiple visual domains, and demonstrate that across all domains \methodname finds abstractions that dramatically simplify the input datasets by discovering meaningful parametric and structural relationships.
With respect to an objective function that tracks how well the input dataset has been abstracted, we find that \methodname significantly outperforms ShapeMOD (even when given access to our wake phase) and DreamCoder (which fails to converge without a curriculum of tasks).
In a series of ablation experiments, we justify the design decisions of our method, and demonstrate the importance of our conditional rewrite scheme and bottom-up recognition network.
Finally, we investigate combining our approach with methods that automatically convert 3D shapes into primitives in an unsupervised fashion, allowing us to discover programs and abstraction functions directly from `in the wild' 3D meshes \cite{shapenet2015}.
In this setting, we observe \methodname still discovers interesting, high-level abstractions, even over noisy, inconsistent primitive decompositions. 

 In summary, our contributions are:
\begin{enumerate}[(i)]
    \denselist
    \item \methodname, a method that learns to infer visual programs that use automatically-discovered abstractions
    to explain and simplify a collection of shapes represented with unstructured primitives;
    \item a recognition network capable of inferring visual programs that use discovered abstractions, even without access to a task curriculum; and 
    \item a refactor operation that augments e-graphs with a conditional rewriting scheme to identify matches on complex parametric relationships in a tractable manner.
\end{enumerate}

Our code is available at github.com/rkjones4/ShapeCoder/ .

\section{Related Work}

Our method is related to a host of prior work that relate in different ways to visual programs: abstraction discovery for non-visual programs, visual program induction (VPI), and visual program generation.
We first provide an overview of these areas, and then end this section with a detailed discussion of the two most related works, DreamCoder~\cite{DreamCoder} and ShapeMOD~\cite{jones2021shapeMOD}.

\paragraph{Program abstraction.}
Several prior methods aim to discover abstractions in context-free languages, where only a reduced set of relations between primitives or sub-programs can be modeled, in the context of fa\c{c}ade grammars~\cite{FacadeInduction} or more general grammar types~\cite{BayesianGrammarInduction, BayesianProgramMerging, ProcmodLearn}. 
Abstraction discovery for more general sets of programs has been explored in the Exploration-Compression algorithm~\cite{ExplorationCompression} and more recently in DreamCoder~\cite{DreamCoder}. Similar to our approach, these methods find abstractions in multiple rounds that alternate between program induction, where programs for a given set of problems are discovered, and abstraction discovery, where discovered programs are examined to find recurring patterns. We discuss DreamCoder in more detail at the end of this section. 
\rev{ 
Recently, improvements have been proposed for the abstraction step of DreamCoder's algorithm \cite{topdownlib}.
Most relevant to our work, Babble \cite{babble}, also uses e-graphs~\cite{EqualitySaturation} to identify abstraction applications, but has no special mechanism for handling rewriters with complex parametric expressions, which allows \methodname to scale to complex 3D visual domains. 
Babble employs anti-unification over e-graphs to propose abstractions, and it would be interesting to consider how this scheme could be extended to work with our e-graph formulation, where we explicitly avoid expanding parametric operations into e-nodes.}
A related problem is to discover common patterns in a \emph{single} program, as opposed to a set of programs. This has been explored for L-Systems~\cite{guo2020inverse} or CAD programs~\cite{Szalinksi, CarpentryCompiler}.

\paragraph{Visual program induction.}
Inferring a visual program that reconstructs a given target is a long-standing problem in computer graphics. Early approaches focused on vegetation~\cite{stava2014inverse, xu2015procedural}, fa\c{c}ades~\cite{muller2007image, wu2013inverse}, and urban landscapes~\cite{vanegas2012inverse, demir2014proceduralization}. We refer to~\cite{aliaga2016inverseproc} for a more complete overview.

In more recent work, neural components are typically employed in key parts of the method.
Some of these approaches require an existing program structure to be available and only estimate the parameters of the program to match a given target 3D shape~\cite{MB:2021:DAGA, pearl2022geocode} or 2D material~\cite{Shi2020:MATch}.
Other approaches aim to jointly infer both program parameters and program structure. Visual programming domains range from commonly used program types, such as CSG construction sequences~\cite{sharma2018csgnet, du2018inversecsg, kania2020ucsgnet, ren2021csg, Yu_2022_CVPR, ren2022extrude}, CAD Modelling Sequences~\cite{Li:2020:Sketch2CAD, Ganin2021ComputerAidedDA, seff2021vitruvion, zoneGraphs, Li:2022:Free2CAD}, SVG shapes~\cite{reddy2021im2vec,reddy2021multi}, and L-Systems~\cite{guo2020inverse}, to custom program domains, like primitive declarations with loops and conditionals in 2D ~\cite{ellis2018learning} \rev{and 3D} \cite{tian2019learning}, geometry instancing with linear transformations~\cite{deng2022unsupervised} and masked procedural noise models for materials~\cite{hu2022inverse}. A few methods also propose inference methods that apply to diverse types of programs~\cite{ellis2019repl, jones2022PLAD}. Most related to our program domain are ShapeAssembly~\cite{jones2020shapeAssembly} and ShapeMOD~\cite{jones2021shapeMOD}, which output programs that arrange cuboid primitives.
All of these methods, excluding ShapeMOD, assume a DSL with a complete set of operators. We discuss ShapeMOD separately at the end of this section.

\paragraph{Visual program generation.}
Several deep generative models have been proposed to generate visual programs. MatFormer~\cite{guerrero2022matformer} generates node graphs for materials, several methods propose generative models for SVG images~\cite{reddy2021im2vec, DeepSVG}, CAD sketches~\cite{para2021sketchgen, seff2021vitruvion, Ganin2021ComputerAidedDA}, and 3D CAD Modelling sequences~\cite{DeepCAD, xu2022skexgen,Li:2022:Free2CAD}. The ShapeAssembly~\cite{jones2020shapeAssembly} and ShapeMOD~\cite{jones2021shapeMOD} methods mentioned above can also be used as generative models. Similar to methods for visual program induction, all of these methods, except for ShapeMOD, require a DSL with a full set of operators.

\paragraph{DreamCoder} \revf{This work proposes} a system that jointly discovers abstractions and performs program induction over arbitrary functional programming languages \cite{DreamCoder}. 
At its core DreamCoder uses three phases to perform this hard task.
A dream phase samples random programs from a library (optionally augmented with abstractions).
A wake phase trains a recognition network to infer programs based on the dream samples. 
An abstraction phase looks over a corpus of returned programs from the wake phase, and proposes and integrates abstractions that improve an objective function.
The objective function trade-offs program likelihood under the library with the complexity of the library.

While DreamCoder's generality \revf{allows it to effectively scale across a wide-variety of program inference tasks}, its abstractions are purely structural, treating real-valued program components as discretizations. 
This means that it is not well-suited for shapes (or other visual domains) where ideally abstractions would capture both complex parametric and structural relationships.
Another challenge of applying DreamCoder to shape programs is that its iterative procedure is reliant on a curriculum to solve tasks: all of its stages (dreaming, waking, abstraction) rely on the assumption that solutions to at least some of the input tasks have a high probability under the current library functions.
When the input tasks form a curriculum (e.g. some tasks are very easy to solve under the base DSL), then this procedure works very nicely, gradually discovering more and more abstractions that allow it to solve increasingly complex VPI tasks.
Unfortunately, when this curriculum assumption is broken, DreamCoder can fail to discover any programs or abstractions for a given domain.
Based on these properties, we ran investigations of how DreamCoder fairs on a simple grammar with parametric relationships, and found that it wasn't able to discover the kinds of abstractions that \methodname is able to find. 
We provide details in the supplemental material.

\paragraph{ShapeMOD}
In contrast to DreamCoder, ShapeMOD is a system designed for visual datasets, like shape programs. 
It has been shown to discover abstractions that extract out meaningful relationships in terms of both parametric expressions and program structure. 
Yet, it does not solve the problem completely. 
ShapeMOD is able to find these abstractions under fairly stringent input assumptions: it requires a collection of imperative programs as input, where all possible valid line reorderings are known. 
In fact, as its intractable to reason over all line reorderings that would lead to the same semantic output, heuristics were employed to limit the orders to a very small set. 
Applying these heuristics required access to a hierarchical semantic segmentation, which allowed sub-parts to be treated as independent sub-programs.
ShapeMOD's integration and proposal stages (analogous to DreamCoder's abstraction phase), relied on these limited program reorderings to both discover candidate macros, and identify when those macros could be applied.

\methodname shares the same goals as ShapeMOD, but aims to discover useful abstractions while making much weaker assumptions: it does not assume access to ground-truth programs, canonical line-orderings, or hierarchy decompositions.
Instead \methodname takes in a dataset where each shape is expressed as an unordered set of primitives.
Discovering abstractions under these assumptions requires both developing logic to infer programs that explain the input shapes, along with extending the abstraction phase so that it is able to reason over arbitrary reorderings of the inferred programs.
We solve the latter problem through the use of e-graphs and a conditional rewrite scheme.

\section{Overview}
\label{sec:met_overview}

\newcommand{\library}{\ensuremath{\mathcal{L}}\xspace}
\newcommand{\function}{\ensuremath{f}\xspace}
\newcommand{\dataset}{\ensuremath{\mathcal{D}}\xspace}
\newcommand{\datapoint}{\ensuremath{d}\xspace}
\newcommand{\objective}{\ensuremath{\mathcal{F}}\xspace}
\newcommand{\program}{\ensuremath{p}\xspace}
\newcommand{\programs}{\ensuremath{\mathcal{P}}\xspace}
\newcommand{\expr}{\ensuremath{e}\xspace}
\newcommand{\abstraction}{\textit{a}\xspace}

\newcommand{\newlibrary}{\ensuremath{\mathcal{L}}'\xspace}
\newcommand{\newprograms}{\ensuremath{\mathcal{P}}'\xspace}
\newcommand{\bestprogram}{\ensuremath{p^*}\xspace}

\newcommand{\types}{T\xspace}
\newcommand{\type}{$\tau$\xspace}
\newcommand{\weight}{$\lambda$\xspace}
\newcommand{\progweight}{$\omega$\xspace}
\newcommand{\dreamnum}{$N_D$\xspace}
\newcommand{\absnum}{$N_A$\xspace}

\begin{figure}[t]
  \includegraphics[width=\linewidth]{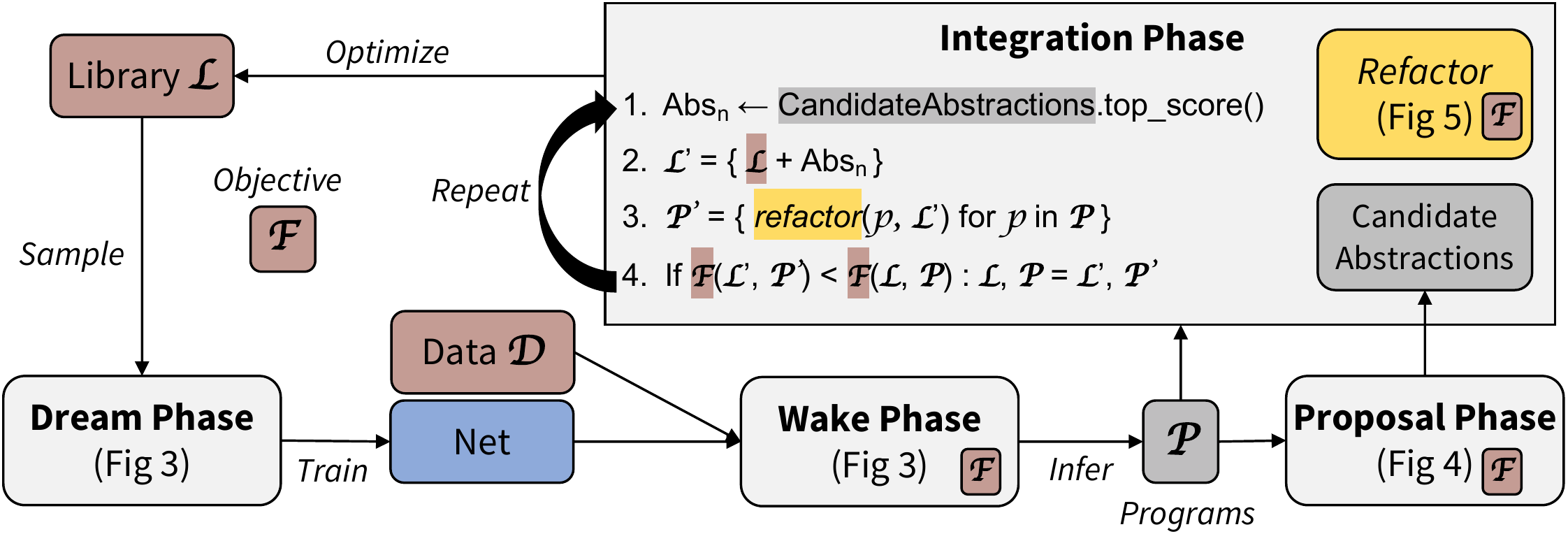}
\caption{ 
\rev{
\textbf{Overview.}
\methodname consumes an initial library~\library, an objective ~\objective, and a dataset of shapes~\dataset (brown boxes).
Each round of the algorithm iterates through a series of phases that
progressively add abstractions into \library to improve \objective. 
A \textbf{dream} phase trains a recognition network by sampling from~\library.
A \textbf{wake} phase infers programs for shapes in ~\dataset.
A \textbf{proposal} phase produces candidate abstractions.
An \textbf{integration} phase uses a \textit{refactor} operation to decide when these abstractions should be added into \library.
}}
\label{fig:overview}
\end{figure}

\methodname  automatically discovers a library of abstraction functions tailored for an input dataset of shapes.
It takes the following as input: 
a library \library describing a functional domain-specific language,
a dataset of shapes \dataset, 
and an objective function \objective. 
Each \datapoint $\in$ \dataset is represented as a collection of unstructured primitives, and we assume that there exists some program expansion of \library, \program, such that executing \program would recreate \datapoint. 

\methodname's goal is to minimize \objective  (Section \ref{sec:met_obj}), which expresses a trade-off between how well-suited \library is for \dataset (program complexity) and how many abstractions functions have been added to \library (library complexity).
We break this task into multiple steps that each tackle a tractable sub-problem.
We depict the distinct phases of \methodname in Figure \ref{fig:overview}.
The \textit{dream} phase (Section \ref{sec:met_dream}) samples scenes from \library to train a program recognition network.
The \textit{wake} phase (Section \ref{sec:met_wake}) uses this network to infer programs \programs that recreate shapes in \dataset .
The \textit{proposal} phase (Section \ref{sec:met_proposal}) consumes \programs as input, and generates candidate abstraction functions.
Finally, the \textit{integration} phase (Section \ref{sec:met_integration}) considers proposed candidate abstractions and finds modified versions of \library to improve \objective, which can be passed in to a subsequent \textit{dream} phase.
Of note, the integration phase uses a \textit{refactor} function (Section \ref{sec:met_refactor}) to find minimal cost equivalent programs under different libraries in a tractable manner through use of e-graphs and a novel conditional rewriting scheme.

In the following sections, we walk-through these various stages, where examples in the text and figures use programs from a toy 2D grammar for rectilinear shapes (Appendix \ref{sec:apndx_grammar}). 
Further implementation details are provided in Appendix \ref{sec:apndx_impl_dets}.

\subsection{Optimization Objective \objective}
\label{sec:met_obj}

\methodname's objective function \objective takes in two arguments: a library~\library and a collection of programs from \library  that correspond with a shape dataset \dataset.
~\objective measures the trade-off between two competing terms: the complexity of \library and \programs.

The complexity of each \program $\in$ \programs is computed according to Occam's razor: all else equal, shorter programs are better.
We compute program length with a weighted sum of program tokens: if \library has token types \types (e.g. booleans, floats, etc.), we allow users to specify a weight~\weight for each \type $\in$ \types. Further, \methodname employs a geometric error function, \textit{err}, that compares the executed geometry of each~\program~$\in$~\programs against its corresponding shape, \datapoint $\in$ \dataset. 
If \textit{err}(\program, \datapoint) returns a value above a user-defined threshold, \objective returns $\infty$.
Otherwise, the error is added into \objective with weight $\lambda_{e}$.

Library complexity can be measured by tracking the number of functions that \library contains.
\methodname allows users to specify a function weighting scheme, \progweight. 
\progweight consumes a function \function from \library and returns a value in the range (0, $\infty$). 
Lower \progweight values make it easier to add \function into \library.
As an example, we find it useful to increase the \progweight of \function according to the number of input parameters \function consumes, as this often indicates an overly general pattern.

With this machinery, where \type(\program) expresses the number of tokens in \program that have type \type , we can express \methodname's objective as:
\begin{equation*}
\small
    \objective(\library,\programs) = \frac{1}{|\programs|} \left(\sum_{\program \in \programs} \left( \sum_{\tau \in T} \lambda_\tau * \tau(\program) \right) + \lambda_{e} * err(\program, \datapoint) \right) + \sum_{\function \in \library} \omega(\function)
    ~~.
\end{equation*}

\begin{figure*}[t!]
 \includegraphics[width=\linewidth]{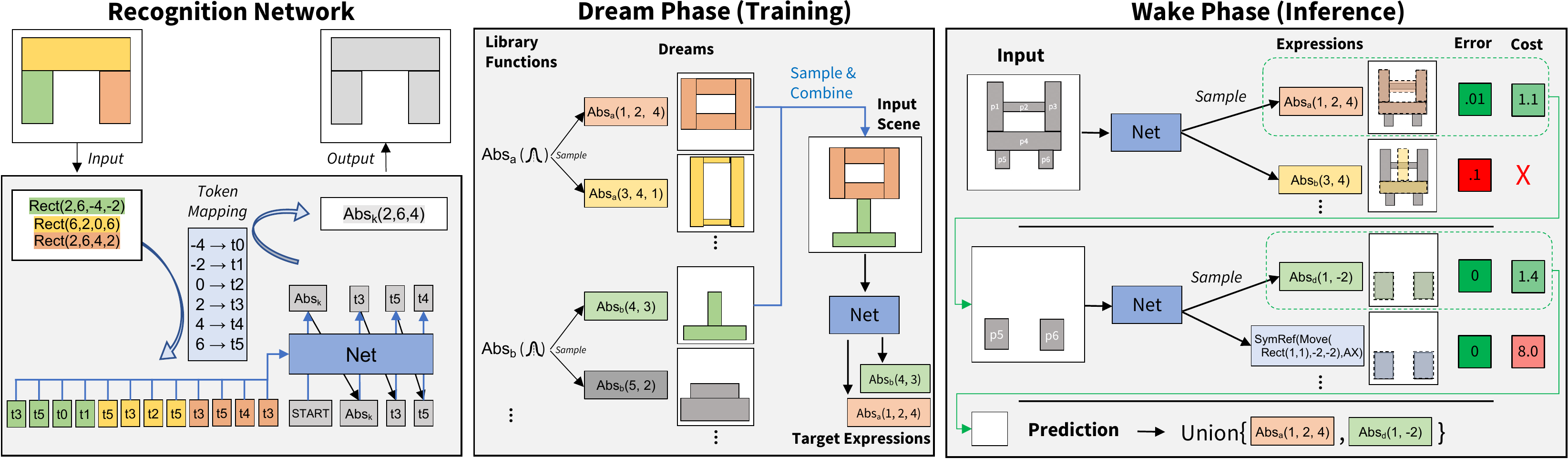}
\caption{\rev{
\textbf{Dream and Wake Phases.} \textit{(Left)} \methodname's recognition network is a Transformer decoder that attends over tokenized input primitives and autoregressively predicts functions and parameterizations. \textit{(Middle)} The dream phase trains the recognition network by sampling expressions from library functions, which are randomly combined together to form (input, target) training pairs. \textit{(Right)} The wake phase uses the recognition network to find programs that explain input shapes. In a series of iterative steps, it samples expressions, chooses the expression that achieves the best cost, and removes covered primitives from the input canvas, until the canvas is empty. 
}}
\label{fig:net}
\end{figure*}

\section{Inferring Visual Programs}
\label{sec:met_infer}

While \methodname consumes a shape dataset \dataset  as input, it doesn't know what programs \programs from a given library version \library can best represent \datapoint $\in$ \dataset.
To solve this problem, \methodname uses a program recognition network (Section \ref{sec:met_arch}), trained on randomly sampled programs from \library (dream phase, Section \ref{sec:met_dream}), to infer \programs that minimize~\objective (wake phase, Section \ref{sec:met_wake}). 

To simplify this search, our recognition network learns to infer partial solutions: expressions from \library that recreate a subset of input primitives.
Found expressions are then combined together to form a complete program that explains an input scene.
This framing requires that \library contains a combinator operation (e.g. \texttt{Union}).
To ensure that our search procedure never fails to find \textit{some} solution, we assume access to an analytical procedure for finding expressions in \library that can recreate any primitive in \datapoint  
(e.g. any cuboid can be represented with a scale, rotation, and translation sequence).

\subsection{Recognition Network}
\label{sec:met_arch}

We depict \methodname's recognition network on the left side of Figure \ref{fig:net}. 
The recognition network consumes a scene of geometric primitives as input, and aims to output an expression from \library that corresponds with a subset of the input primitives.  
We implement this network as a Transformer \cite{att_is_all} decoder that autoregressively predicts a sequence of tokens from \library.
The network is conditioned (through causal-masking) on an encoding of the input primitives: if $M$ primitives are each represented with $K$ parameters, the network attends over $K \times M$ conditioning tokens ($M=3$ and $K=4$ in the figure example).  
To convert expressions into token sequences, discrete elements of \library are given a unique index. 
To tokenize real-valued parameters, we employ a simple mapping procedure:
for a given input scene, we take all real values in the primitive parameterizations, bin them through rounding (to 2 decimal places), and sort them to produce a token mapping (light-blue box).
This mapping is used to form the conditioning tokens, and converts network predictions back into real values.

\subsection{Dream Phase}
\label{sec:met_dream}

The dream phase trains the recognition network by randomly sampling example scenes from \library. 
We show this process in the middle box of Figure \ref{fig:net}.
To begin the dream phase, for each function~\function~$\in$~\library, \methodname creates \dreamnum  number of dreams for~\function .
Dreams are generated by sampling random instantiations of each parameter slot of~\function.
Rejection sampling is employed to avoid dreams that create bad geometry by checking easy to enforce properties (geometry outside scene bounds, primitives with negative dimensions, primitives wholly contained by other primitive, etc.). 

However, as shapes in \dataset often contain scenes best explained by more than one function, its not enough to train on function-specific dreams directly. 
We solve this issue with composite scenes formed by sampling function-specific dreams and combining their output primitives together (blue arrow).
If a composite scene was formed by combining $K$ sampled dreams, then we can derive $K$ paired training examples for the recognition network: the input to the network will be the composite scene, and each of the $K$ sampled dreams would be a target output. 
For instance, given the input scene with orange and green primitives in Figure \ref{fig:net}, we would train the network to predict both the green and orange expression sequences (i.e. there is a one-to-many mapping). 
Once this paired data has been assembled, by ensuring that each \function $\in$ \library appears in at least \dreamnum  target sequences, the recognition network can be trained in a supervised fashion with maximum likelihood updates.

\subsection{Wake Phase}
\label{sec:met_wake}

The wake phase takes an input shape \datapoint  and aims to infer a program~\program that minimizes \objective using the recognition network. 
We depict this process on the right side of Figure \ref{fig:net}.

To begin, the scene is initialized to contain the primitives of \datapoint . 
Then the wake phase performs the following steps in an iterative fashion. 
First the input scene is used to condition the recognition network, which samples a large set of expressions from \library according to its output probabilities, up to a timeout (1 second).
For every sampled expression, \expr, we record its cost: the program complexity of \expr under \objective, normalized by the number primitives it explains. 
Note that if \expr does not recreate a subset of primitives in the input scene, it will have a high geometric error, and \objective will return $\infty$ (red X in figure). 
The wake phase chooses the lowest cost $\expr^*$ (dotted green lines), and removes all primitives it covers from the input scene, which is then fed back into the recognition network.
These steps are repeated until the canvas is empty. 
Once this condition is met, the final program \program explaining \datapoint is formed by applying the combinator operation in \library over each $\expr^*$ (e.g. the \texttt{Union} of the orange and green expressions in the bottom-row). 
For every input scene, the `naive' expression for a single primitive under \library is added to the sampled set of expressions, so that a valid solution is guaranteed to be found.

During each \methodname round, the wake phase uses the recognition network to infer a set of programs that explain \dataset. 
But should we treat these predictions independently?
One option is to clear all program entries in \programs before every wake phase.
However, this would cause \methodname to `forget' good solutions discovered in previous rounds. 
Instead, we use the following approach: 
for round~$r$,~$r>0$, if \programs contains previously discovered programs, and $\programs_r$ contain programs discovered in round $r$'s wake phase, then we set each entry of \programs to be the result of \textit{combine}(\program, $\program_r$), where \textit{combine} performs a greedy replacement search to optimize \objective.

\begin{figure*}[t!]
   \includegraphics[width=\linewidth]{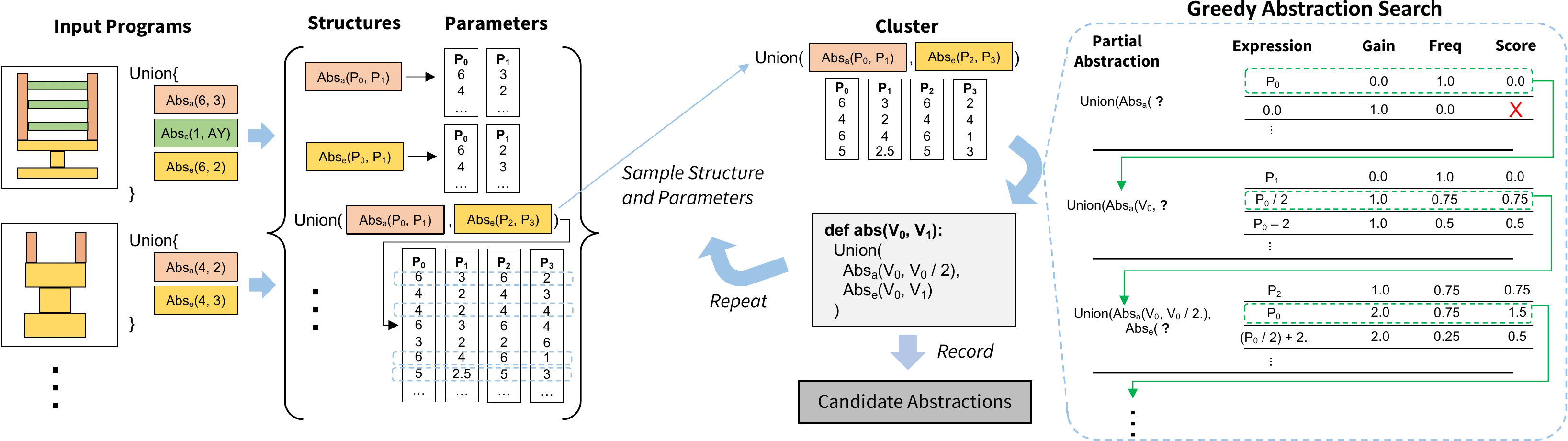}
\caption{\rev{\textbf{Proposal Phase.} The proposal phase consumes a collection of programs and outputs a set of candidate abstractions. 
First, possible structures and their parameterizations are recorded from the input programs. Then clusters are formed by sampling a structure and a subset of parameterizations. For each cluster, a greedy abstraction search generates a possible abstraction, which is recorded.}}
\label{fig:proposal}
\end{figure*}

\section{Proposing and Integrating Abstractions}
\label{sec:met_abstraction}

Together, the dream and wake phases train and use a recognition network to infer a set of programs \programs that explain the shapes of the input dataset \dataset.
The proposal phase (Section \ref{sec:met_proposal}) reasons over \programs to suggest candidate abstractions functions, used by the integration phase (Section \ref{sec:met_integration}) to find library variants that improve \objective.

\subsection{Proposal Phase}
\label{sec:met_proposal}

The goal of the proposal phase is to search over \programs for abstraction functions that would improve \objective if added into \library. 
As this search is computationally intractable to solve globally, \methodname's proposal phase instead solves more tractable sub-problems (subsets of~\programs), and aggregates local solutions. Figure \ref{fig:proposal} outlines this process. 

\paragraph{Identifying Structures and Parameters.} 
As \library is a functional language, generating an abstraction \abstraction requires two steps: deciding the structure of \abstraction(what are its sub-functions) and deciding how \abstraction is parameterized (what input does \abstraction take, and how are those mapped to its sub-functions).
What structures should we consider for possible abstractions? 
Each program \program $\in$ \programs is found in the wake phase by combining expressions that solve sub-tasks, so \program will have no consistent or canonical ordering.
Therefore, we would like to factor out expression ordering by considering structural variants over any possible function reordering of each \program $\in$ \programs. 
However, as the general solution is intractable, we instead consider a limited set of potential abstraction structures: singleton and paired combinations of sub-expressions found in \programs.
We record all such observed structures as keys and how those structures were parameterized as values (see bracketed data structure in figure). 
We additionally find it useful to apply a simple filtering step that removes infrequently observed structures in \programs from this mapping (seen in less than 5\% of \programs).
 
\paragraph{Cluster Sampling and Search.} 
Once potential structures and their observed parameterizations have been recorded, the proposal phase begins an iterative process.
To convert the global problem into a local one, a random structure and a subset of its parameterizations are sampled to form a cluster.  
Then a greedy search is run over this cluster to find an abstraction \abstraction that would optimize~\objective. 
The generated function is recorded into a candidate abstraction data structure that keeps track of a coverage set of \program $\in$ \programs that could be simplified through applications of \abstraction.
This procedure is repeated many times, and coverage sets are expanded whenever the candidate abstraction data structure receives a previously observed abstraction.

\paragraph{Greedy Abstraction Search}
We employ a greedy search to find an abstraction \abstraction for a given cluster (right side Figure \ref{fig:proposal})
This search is guided by a \textit{score} function that provides a heuristic estimate of how \abstraction would improve \objective if it were added into \library.
The \textit{score} of \abstraction is a product of two terms: the \textit{frequency} and the \textit{gain}.
The \textit{frequency} (Freq column in figure) is the percentage of instances in the cluster that \abstraction could recreate (with the correct parameterization). 
The \textit{gain} tracks the number of parameters removed from a program \program, whenever \program could be rewritten with \abstraction, denoted as $\program_\abstraction$.
For instance, the proposed abstraction in Figure \ref{fig:proposal} would remove two float-typed parameters whenever it could be applied, corresponding with slots $P_1$ and $P_3$ in the cluster. Using the weighting from \objective(Section \ref{sec:met_obj}), we have:
\begin{equation*}
    \mathrm{\textit{gain(\abstraction)}} = \sum_{\tau \in T} \lambda_\tau * (\tau(\program) - \tau(\program_\abstraction))  
    ~~.
\end{equation*}
The function sequence in the proposed abstraction is determined by the structure of the sampled cluster, but how should we fill in the parameter slots?
For each slot, we consider a set of possible expressions, calculate the \textit{score} of each option, and add the expression with the highest score into the partial abstraction. 
If the \textit{frequency} is ever zero, then the \textit{score} is voided. 
For float-typed parameter slots, \methodname produces expressions by iterating over a preference ordering of possible parametric relationships. 
For discrete-typed parameter slots, a previously instantiated parameter can be reused, or a static value can be assigned. 
This search always includes defining a new free parameter (e.g. using the parameterization in the sampled cluster) as an option (depicted as the top-row of each step).

\subsection{Integration Phase}
\label{sec:met_integration}

\begin{figure*}[t!]
  \includegraphics[width=\linewidth]{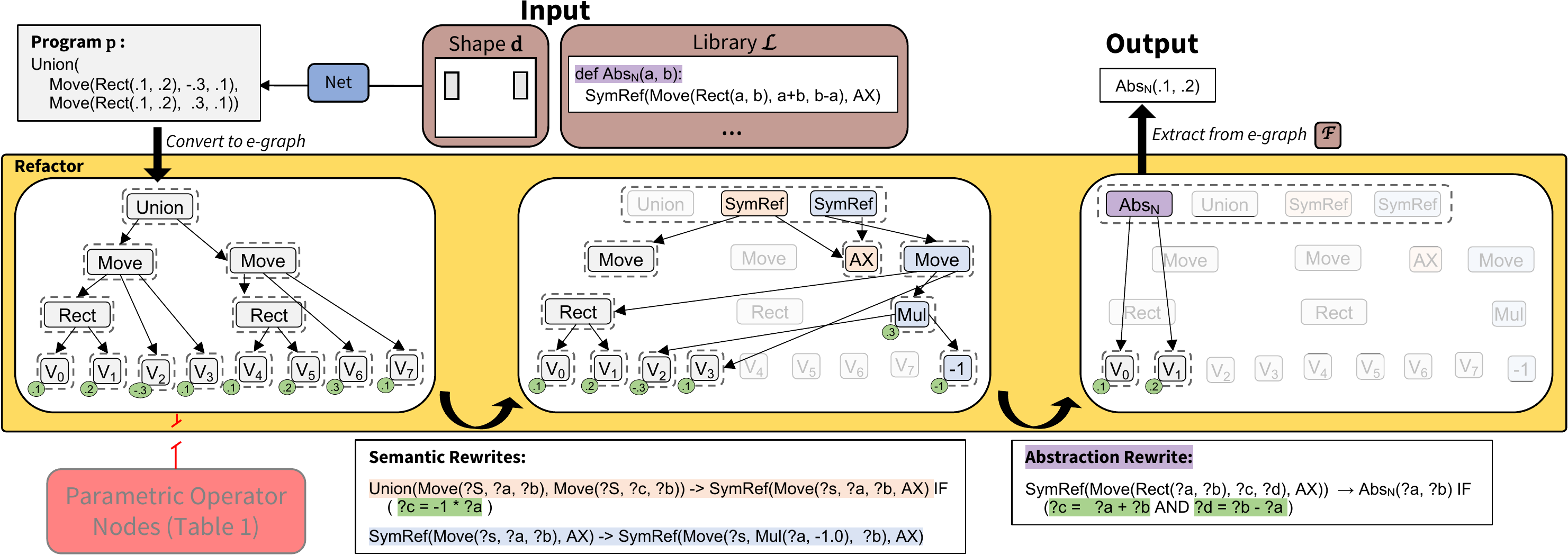}
\caption{\rev{\textbf{Refactor.} The refactor operation uses e-graphs to identify when abstractions can be applied. 
Input programs are converted into e-graphs, which are expanded with semantic and library-specific rewrites to uncover lower-cost equivalent expressions that can be extracted.
We develop a conditional rewrite scheme that reasons over parametric relationships (green highlights) without adding excessive e-nodes for parametric operators (red box).
}}

\label{fig:spad_egraph}
\end{figure*}

The integration phase takes in a library \library, a set of programs \programs, and candidate abstractions from the proposal phase.
It searches for modified version of \library that can be used to refactor \programs to improve \objective.
The \textit{refactor} operation (Section \ref{sec:met_refactor}) uses e-graphs to efficiently search for minimal cost equivalent programs under different \library variants.

The integration phase begins by first recording the starting objective value: \objective(\library, \programs). 
It then iterates through a series of steps in an attempt to greedily improve this value.
First, a new library variant~\newlibrary is formed by sampling a candidate abstraction and adding it into~\library.
The abstraction with the top \textit{score} value is chosen, where the notion of \textit{frequency} is generalized from clusters to all of \programs. 
Then a new program set, \newprograms, is formed by applying the \textit{refactor} operation over each \program $\in$ \programs under \newlibrary.
Finally, if \objective(\newlibrary, \newprograms) is better~than~\objective~(\library,~\programs), both~\library and~\programs are replaced with their modified versions.

Evaluating a modified library \newlibrary is expensive, as it requires running the \textit{refactor} operation for every \program $\in$ \programs, so we usually consider a small number, \absnum, of top-ranked candidate abstractions during each integration phase.
To keep the \textit{score} heuristic as accurate as possible, whenever \newlibrary, that added \abstraction to \library, improves \objective, we check which \program $\in$ \programs contributed to the \textit{frequency} of \abstraction and discount the \textit{frequency} of other abstractions that overlapped on the covered set.

Beyond this greedy search, two other forms of library variants are also considered during the integration phase.
Whenever adding~\abstraction to~\library does not improve \objective, we compute the set of functions whose frequency between \programs and \newprograms decreased significantly; call this set~$\function_{dec}$. 
We then consider $\library_{dec}$ = \{ \library + \abstraction~-~$\function_{dec}$ \} as a library variant.
This procedure allows the greedy integration search to avoid a local minima where \abstraction would not be added to \library because similar (but worse) functions already exist in \library.
In addition, to finish the integration phase, we consider library variants where each \function $\in$ \library is removed one at time.
In all comparisons, the library variant becomes the new default if it improves the objective function.
At the end of the integration phase, the \library that achieved the best \objective score is then passed into the subsequent dream phase to begin a new \methodname round.

\section{Refactoring Programs with E-Graphs}
\label{sec:met_refactor}

\methodname's integration phase evaluates how library variants can be used to compactly represent \programs  but how does it know when abstractions can be applied?
For this task, we use the \textit{refactor} operation: it takes as input a program, \program, and aims to find \bestprogram, an equivalent program to \program that minimizes \objective. 
This is a hard search problem, which we make tractable through the use of e-graphs~\cite{eq_sat_tate} and a conditional rewriting scheme. 
In the rest of this section, we provide a quick background on e-graphs, and walk-through their role in \textit{refactor} with a running example, depicted in Figure \ref{fig:spad_egraph}.

\paragraph{Background on e-graphs.}
E-graphs are a specialized data structure capable of efficiently representing a large set of equivalent programs.
We show an example e-graph in the left call-out of the figure. 
E-graphs are made up of e-nodes (solid boxes) and e-classes (dotted boxes)
Each e-node is associated with a term from \library and has a pointer (arrows) to e-class children, if that term is a function. 
Each e-class contains a set of equivalent e-nodes.
The root of the e-graph is the e-class that contains the e-node associated with the outermost operator in the input expression (\texttt{Union} in the figure).

This representation becomes useful when it is combined with \textit{rewrite} rules. 
Rewrite rules are domain-specific, pattern matching program transformations that maintain semantic equivalence.
For instance, for any ?a and ?b: \texttt{Union}(?a, ?b) is equivalent to \texttt{Union}~(?b,~?a).
E-graphs are expanded by iteratively applying rewrite rules to create new e-classes and new e-nodes.
These newly created constructs reference existing e-class and e-nodes, allowing the e-graph to represent a large set of equivalent programs in a space-efficient manner.
Importantly, e-graphs also provide support for quickly finding minimal cost rewritten versions of a starting expression, by running a greedy recursive algorithm starting at the root e-class.

\paragraph{Refactor Operation.} The refactor operation consumes an input program \program from the wake phase.
First, it converts \program  into an e-graph, as depicted in the left call-out of Figure \ref{fig:spad_egraph}. 
In this step, each float-typed token is replaced with an independent variable ($V_0$ to $V_7$). 

The operation also consumes a library \library as input.
It uses \library to source two types of rewrite operations.
\textit{Semantic} rewrites express domain-knowledge over base DSL functions and are provided as part of the language definition.
For instance, the blue rewrite expresses the following logic: a sub-expression ?s moved to \textit{xy} position~(?a,~?b) and reflected over the X axis is equivalent to moving ?s to \textit{xy}~position~(-1~$\times$~?a,~?b) and reflecting it over the X axis. 
\textit{Abstraction} rewrites correspond with the abstractions in \library, where rewrites express the conditions that need to be met in order for the abstraction to be applied.
For instance, $Abs_N$ (top-middle) in the input library creates the purple highlighted abstraction rewrite (lower-right).

\methodname expands the e-graph by iteratively applying these rewrite operators. 
In the middle-frame, the orange rewrite first introduces a new \texttt{AX} e-node into a new e-class and a new \texttt{SymRef} e-node into the root e-class. 
Following this, the blue rewrite can be applied, matching on the orange e-nodes, to add the blue highlighted e-nodes.
At this point, the purple abstraction rewrite can be applied, and a new $Abs_N$ e-node is added into the root e-class.
The refactor operation will continue expanding the e-graph until it is saturated (nothing can be added) or a timeout is reached.

Once the rewrites have expanded the e-graph, we can run an extraction procedure on the root e-class to find the minimum cost expression \bestprogram in the e-graph equivalent to the starting program~\program. 
In this example, \bestprogram will be equal to $Abs_N$($V_0$, $V_1$), which we can rewrite to $Abs_N$(.1, .2) using the reverse of the parameter mapping we used to convert the initial program into an e-graph.

\paragraph{Conditional Rewrite Scheme.}
The above explanation is complete up to one critical step: how do know when rewrites can be applied?
E-graphs typically search for structural pattern-based matches, and some semantic rewrites can be included in this framework (e.g. the blue rewrite). 
However, other rewrites, such as the purple abstraction rewrite, require both structural and parametric matches.
For instance, the structural matching requirement to apply $Abs_N$ would be finding some sub-graph of e-classes that matches the pattern of: 
\noindent
\texttt{SymRef(Move(Rect(?a,?b),?c,?d),AX)}, where ?a through ?d can be filled in with any e-class. Beyond this, applications of $Abs_N$ also require parametric matching with logic expressed in green highlights: the ?c spot must be equal to the sum of the ?a and ?b slots, and the ?d spot must be equal to the ?b slot minus the ?a slot.  

How we can support this type of parametric matching? 
A naive solution would convert parametric constraints into structural ones:

\noindent
\texttt{SymRef(Move(Rect(?a,?b),Add(?a,?b),Sub(?b,?a)),AX)}.
The issue with this approach is that it requires adding e-nodes for parametric operations (e.g. \texttt{Add} or \texttt{Sub}) into the e-graph, before it is known whether or not that e-node will be useful. 
When there are many input parameters ($V_i$'s) this naive solution will blow up the size of the e-graph, making the refactor operation ineffective.
We visualize our choice to avoid this blowup with the disconnected red box in the figure.

\methodname addresses this issue of exploding e-graph size by leveraging a conditional rewrite scheme. 
Conditional rewrites are rewrite operations that first find structural matches but only make a rewrite application if additional checks pass.
In this way, each parametric relationship (green highlights on rewrites) is only evaluated lazily, after a structural match has been identified. 

Concretely, in the working example applying the purple rewrite will find the following matches: ?a with $V_0$, ?b with $V_1$, ?c with \texttt{Mul}~($V_2$,-1), and ?d with $V_3$.  
To check that the parametric relationships hold, we need to know the real value associated with each matched e-class. 
Then to check a relationship such as ?d = ?b - ?a, we can simply compare the difference in values between $V_3$~and~$V_1$~-~$V_0$.
This check does not enforce exact matches, but rather allows the user to specify a maximum error threshold, allowing us to apply \textit{approximately}-equivalent rewrites, which is typically a limitation of e-graphs.

For some e-classes, finding their associated real-values is trivial: for each e-class associated with a float-typed parameter e-node ($V_0$ to $V_7$) we record a mapping between e-class ids and values. 
This procedure is complicated by the fact that some rewrites create new float-typed nodes (e.g. the blue \texttt{Mul} e-class).
We handle this case by dynamically updating the e-class-to-real-value mapping during all rewrite steps (represented with green-highlights on e-classes), which is a constant time operation.
Our conditional rewrite step is just as fast as a non-conditional rewrite step and critically avoids unnecessarily expanding the e-graph with unneeded parametric operator e-nodes.
In sum, conditional rewrites provide a dramatic speedup over the naive approach for the kinds of refactoring problems that \methodname typically reasons over (see Table \ref{tab:rewrite_comp}). 

\begin{table}[]
    \centering    
    \footnotesize
    \caption{Comparing our conditional rewriting scheme against the naive alternative. The conditional scheme is able to quickly saturate the e-graph (time reported in seconds), even for complex input expressions with many parameters. The naive approach times out when the complexity is too high. }
    \begin{tabular}{@{}lccc@{}}
        \toprule
        \textbf{Rewrite Scheme} 
        & 8 params & 16 params & 32 params \\
        \midrule
        \textit{Naive} & .22 & 2.6 & X \\
        \textit{Conditional} & .01 & 0.04 & 2.1 \\
        \bottomrule
    \end{tabular}
    \label{tab:rewrite_comp}
\end{table}
\section{Results and Evaluation}
\label{sec:results}

We run \methodname over distributions of visual shapes represented as collections of unstructured primitives.
We describe these domains in Section \ref{sec:res_domain}.
In Section \ref{sec:res_discovery}, we compare how well the abstractions discovered by \methodname improve the objective function compared to alternative approaches.
Our main comparison is against ShapeMOD~\cite{jones2021shapeMOD}.
In the main text, we do not include comparisons against DreamCoder~\cite{DreamCoder},
as we found it performed poorly on a toy grammar with parametric relationships (see supplemental). 
In Section~\ref{sec:res_qual}, we analyze properties of the discovered abstractions and investigate their generality with a post hoc inference procedure.
In Section \ref{sec:res_ablations}, we run an ablation experiment to investigate the importance of various algorithm components.
In Section \ref{sec:res_unstruct} we show another application of our method: inferring visual programs, that contain abstractions, given only a dataset of 3D meshes as input, where we leverage noisy primitives sourced from a pretrained unsupervised cuboid decomposition approach \cite{yang2021unsupcsa}.
Finally, in Section \ref{sec:res_abs_benefits} we explore how \methodname's discovered abstractions benefit downstream tasks.

\subsection{Experimental Domains}
\label{sec:res_domain}

For the main result section, we consider domains of 3D shapes. 
We provide experimental results over a toy dataset of 2D shapes in the supplemental.
Our experiments use manufactured objects sourced from PartNet \cite{PartNet}, where manual annotations are used to convert each 3D object into an unstructured collection of cuboids, that represent part bounding boxes.
We follow past-work in the 3D shape abstraction discovery space, and run experiments on shapes from the \textit{Chair}, \textit{Table}, and \textit{Storage} categories of PartNet.
We perform the cuboid simplification steps outlined by \cite{jones2020shapeAssembly}, so that our starting primitive set is the same as that used by ShapeMOD, except we remove all hierarchy and canonical ordering information.

The DSL (Appendix \ref{sec:apndx_grammar}) we use for our experiments has 4 low-level operations: (i)~instantiating a primitive (\texttt{Cuboid}); (ii)~moving a shape (\texttt{Move}); (iii)~rotating a shape (\texttt{Rotate}); and (iv)~unioning two shapes together (\texttt{Union}).
We also provide two mid-level symmetry operations in the base DSL, that correspond with (v)~reflectional and (vi)~translational symmetry (\texttt{SymRef} and \texttt{SymTrans}). 
  
\subsection{Discovering Abstractions}
\label{sec:res_discovery}

For each PartNet category, we run \methodname for four rounds over 400 shapes from that category.
\methodname is implemented in Python and Rust, using PyTorch and Egg, an e-graph library \cite{2021-egg}. 
We run \methodname on a machine with a GeForce RTX 3090 Ti GPU and
an Intel i7-11700K CPU, and find that it takes less than 24 hours to finish discovering abstractions for a single category (taking at most 4GB of GPU memory).

\begin{table}[]
    \centering    
    \footnotesize
    \caption{
    Abstraction discovery performance, measured with objective function \objective, for libraries of abstractions discovered by different methods. 
    }
    \begin{tabular}{@{}llcccc@{}}
        \toprule
        \textbf{Category} & \textbf{Method} 
        & \textbf{\objective} $\Downarrow$
        & \textbf{$|\library|$} 
        & \textbf{Num Struct}
        & \textbf{Num Param} 
        \\
        \midrule
        \multirow{4}{*}{\emph{Chair}}
        & Input Prims & 146.0 & 6 & 29 & 61 \\
        & ShapeMOD+Input & 109.0 &  21 & 16 & 46\\
        & ShapeMOD+Wake & 83.0 & 21 & 12 & 36 \\
        & \methodname & \textbf{63.6} & 33 & 10 & 27 \\
        \midrule
       \multirow{4}{*}{\emph{Table}}
        & Input Prims & 125.0 & 6 & 25 & 51\\
        & ShapeMOD+Input & 84.2 & 25 & 11 & 34 \\
        & ShapeMOD+Wake & 69.1 & 17 & 10 & 30\\
        & \methodname &  \textbf{40.9 } & 37 & 8 & 18\\
        \midrule
        \multirow{4}{*}{\emph{Storage}}
        & Input Prims & 154.0 & 6 & 30 & 62\\
        & ShapeMOD+Input & 119.0 & 16 & 20 & 48 \\
        & ShapeMOD+Wake & 103.0 & 10 & 19 & 45 \\
        & \methodname & \textbf{71.3 }& 31 & 11 & 33 \\
        \bottomrule
    \end{tabular}
    \label{tab:compress}
\end{table}

\paragraph{Discovering abstractions that improve our objective}

We report how the abstractions discovered from \methodname impact the objective function we optimize over, in Table \ref{tab:compress}.
From left to right, the columns express the objective function score (\objective, where lower is better), the number of functions that the library contains ($|\library$|), and the average number of operations (\textit{Num Struct}) and parameters (\textit{Num Param}) that are needed to represent the input dataset of shapes using programs that make use of the discovered abstractions. 

The top \textit{Input Prims} row for each category conveys the starting objective function value for \methodname. 
This row reports the cost of using `naive' programs to cover the primitives of the input shapes, where each primitive is rotated, moved, and instantiated, whenever that command would have an effect (e.g. moving zero distance would be ignored).
The final objective function score found by \methodname, in the bottom rows, is dramatically better than this starting point. For \textit{Chairs}, \textit{Tables}, and \textit{Storage}, the starting objective function value drops by 56\%, 67\%, and 53\%, respectively. 
This improvement is achieved by adding abstraction functions (2nd column) that remove degrees of freedom needed to represent the shapes of the input set (3rd and 4th columns).

We also compare how \methodname performs against ShapeMOD in this setting. 
The ShapeMOD algorithm requires a dataset of imperative programs as input, along with the possible ways that the lines of the programs can be ordered.
As we lack ground-truth programs for our problem setting, we compare against two versions of ShapeMOD, that attempt to optimize the same objective function as \methodname:

\begin{itemize}
\denselist
    \item \textit{ShapeMOD+Input:} We take the `naive' programs that can be directly parsed from the input collection of primitives, and provide this as input to ShapeMOD.
    \item \textit{ShapeMOD+Wake:} We take the output from \methodname's first wake phase as the input to ShapeMOD. Note that the only `non-trivial' functions in the library for the first wake phase are the symmetry operations, roughly equivalent to running symmetry detection on the `naive' programs.
\denselist
\end{itemize}
For both program datasets, we have no way of knowing how the various expressions (e.g. sub-shapes combined through \texttt{Union}) should be ordered, so we pass a random subset of all possible valid orderings to ShapeMOD, as without limiting the set of orders ShapeMOD takes prohibitively long to run (see supplemental).

Comparing ShapeMOD variants and \methodname in Table \ref{tab:compress}, it is clear that \methodname finds abstractions that significantly improve the objective function over those found by ShapeMOD.
\rev{While} \methodname's wake phase provides a better starting point than the `naive' programs, \rev{in either case}, the complexity of the input programs is too high for ShapeMOD to handle-well when canonical orderings and hierarchy annotations are absent.

\rev{We also compare \methodname against approaches that operate over single programs, like Szalinski \cite{Szalinksi}. 
Szalinski also uses e-graphs in the context of visual programs, and while its fixed rewrite rules are well-suited for simplifying a single heuristically-inferred CAD program of a mechanical object, we found that these rules did not significantly compress shape programs in our domain: Szalinksi’s rewrites improved our objective function from 146~to~131, for chairs, whereas ShapeCoder reached 63.}

\subsection{Analysis of Discovered Abstractions}
\label{sec:res_qual}

\begin{figure*}[]
  \includegraphics[width=.95\linewidth]{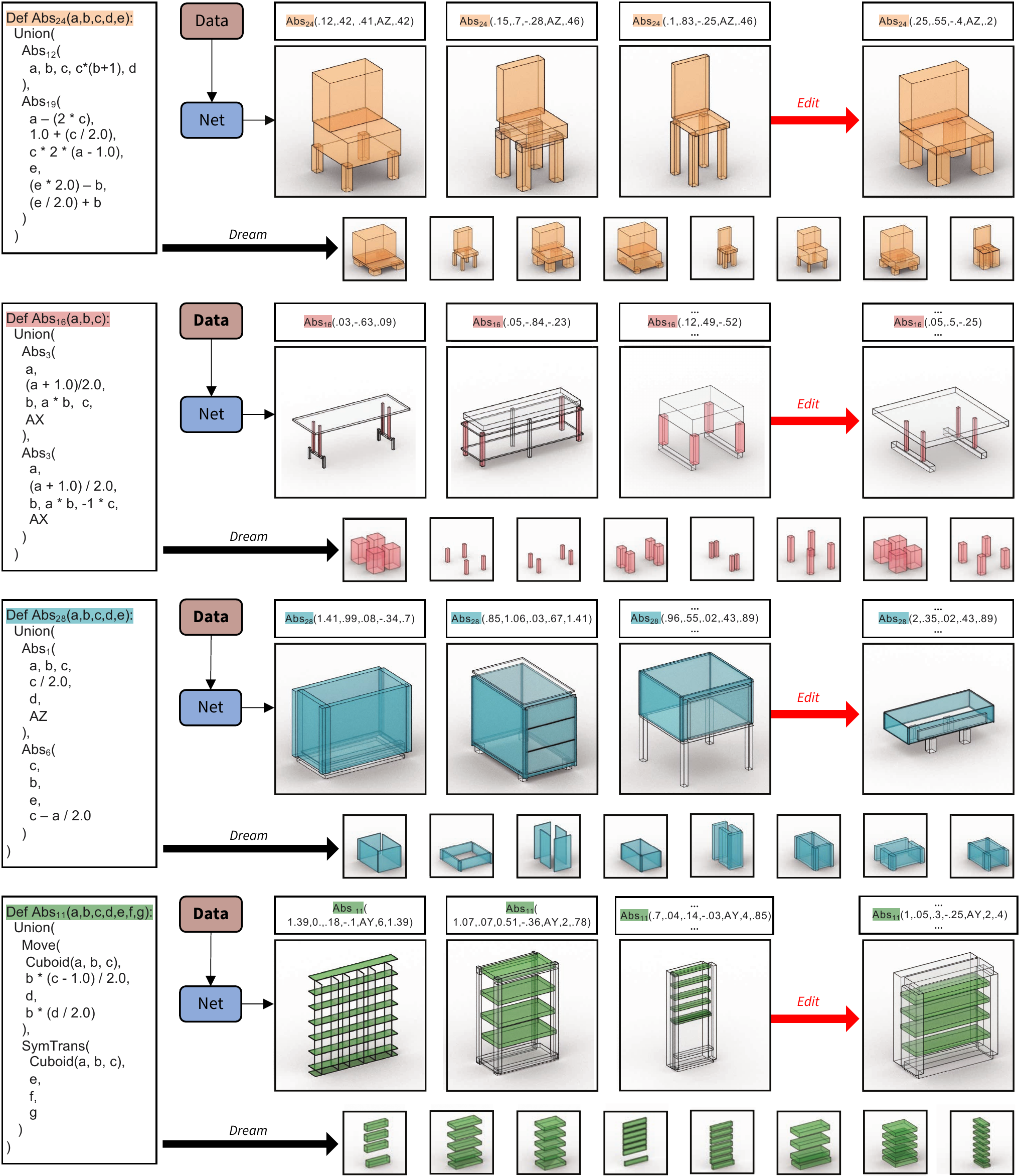}
\caption{Qualitative examples of discovered abstractions. We show one abstraction each for \textit{Chair} and \textit{Table}, and two abstractions for \textit{Storage} furniture. The abstraction code is shown on the left, 
followed by three different usages of the abstraction in our shape dataset discovered by \methodname.
In the right-most column, we manually edit the discovered program to create a new shape. 
Along the bottom, we visualize randomly sampled dreams.
}
\label{fig:qual_abs}
\end{figure*}

We visualize a subset of abstractions discovered by \methodname when run over PartNet shapes in Figure ~\ref{fig:qual_abs}. 
The recognition network learns how to use these abstractions to explain shapes in the input dataset (first three columns).
Programs rewritten with these abstractions can be edited to create new shapes, as we show in the fourth column.
The discovered abstractions contain many desirable properties: they capture diverse geometric expressions and constrain many extraneous degrees of freedom by introducing parametric relationships. 
Abstractions in later rounds of \methodname can reference previously discovered abstractions in sub-function calls, forming a nesting hierarchy of abstractions.
In extreme cases, \methodname can even discover single abstractions that explain entire input shapes, e.g., in the first and third columns of the top-row, a single abstraction function, that consumes five input parameters can output an entire chair when executed.  
Access to these types of abstractions can even be helpful for structural analysis of 3D shapes. 
For instance, the shown abstraction for tables (2nd row) is consistently mapped to the same semantic part (regular table legs), even though the part has a wide range of possible output geometries.
For each abstraction, we also visualize a subset of random parameterizations (i.e. dreams), to give a sense of the possible output space described by each function.

 \paragraph{Post hoc inference}

During the course of abstraction discovery, \methodname finds programs that use abstractions to explain the shapes in its input dataset. 
We investigate if these abstractions can generalize to shapes from the same distribution that were not included in its optimization procedure. 
We leverage \methodname's recognition network to find programs that explain shapes that were not included in the `training' phase of abstraction discovery.
We run the wake phase over these shapes, to find programs that explain the input set of primitives.
These programs are then passed through the \textit{refactor} operation, to see if any of the library rewrites can further improve the program.

\begin{table}[t!]
    \centering    
    \small
    \caption{We measure the generality of the abstractions that \methodname discovers by comparing how well it can compress shapes (objective function~\objective) from a held-out set (Val) with post hoc inference ~(PHI)  compared with the programs it discovers during normal operation (top-row). }
    \begin{tabular}{@{}llcc@{}}
        \toprule
        \textbf{Shape Set} 
        & \textbf{Inference Method}
        & \textbf{\objective} 
        & \textbf{Abs Count} 
        \\
        \midrule
        Train &  \methodname & 63.6 & 4.31 \\
        Train & PHI & 67.5 & 4.67  \\
        Val  & PHI & 70.6 & 4.77 \\
        \bottomrule
    \end{tabular}
    \label{tab:hold_out_quant}
\end{table}

We present the results of this \textit{post hoc inference} (PHI) procedure in Table \ref{tab:hold_out_quant}, for shapes from the \textit{Chair} category of PartNet. 
The top row of this table shows the objective function values, and the average number of abstraction-uses, for the programs that were  iteratively built up during \methodname `training' (e.g., abstraction discovery). 
In the middle row, we take this same set of shapes, `forget' the programs discovered during abstraction discovery, and run the PHI procedure, which aims to infer programs from scratch.
In the last row, we run PHI on validation shapes, never before seen by \methodname.
While doing inference post hoc is slightly worse than iteratively discovering programs over multiple rounds, the difference between running PHI over the `training' shapes and `validation' shapes, is relatively small.
This fact, along with the consistently high-values in the abstraction usage column, indicates that many of the abstractions that \methodname discovers can generalize beyond the dataset of shapes it optimizes over. 

\begin{table}[]
    \centering
    \small
    \caption{\rev{\textit{(Left)} Ablating design decisions of \methodname by tracking objective function improvement (see condition details in Section \ref{sec:res_ablations}). Our default configuration (bottom) performs best. \textit{(Right)} Measuring output execution validity (with Frechet Distance) under increasing perturbations (Noise Level) for programs with, or without, abstractions. Abstractions help keep shapes `in distribution' under parameter edits.  }}
    \begin{minipage}{0.4\linewidth}
    \begin{tabular}{@{}lr@{}}
        \toprule
        \textbf{Condition} 
        & \textbf{\objective} $\Downarrow$ 
        \\
        \midrule
        No Abstraction &  104.9 \\
        Single Iter & 81.6\\
        No Dream+Wake & 99.0\\
        No Semantic Rws & 75.2\\
        No Conditional Rws & 100.0 \\
        No Abs Preferences & 70.7 \\
        \methodname &  \textbf{63.6} \\
        \bottomrule
     \end{tabular}
    \end{minipage}%
     \hspace{2em}
    \begin{minipage}{0.4\linewidth}

 \begin{tabular}{@{}lcc@{}}
        \toprule
        \textbf{Noise Level} & No Abs & With Abs
        \\
        \midrule

        0.1 & \textbf{8} & \textbf{8} \\
        0.2 & 18 & \textbf{13} \\
        0.3 & 40 & \textbf{27} \\
        0.4 & 88 & \textbf{48} \\
        0.5 & 157 & \textbf{84} \\
        
        \bottomrule
     \end{tabular}
 
    \end{minipage}
    \label{table:joint_table}
\end{table}

\begin{figure*}[]
  \includegraphics[width=\linewidth]{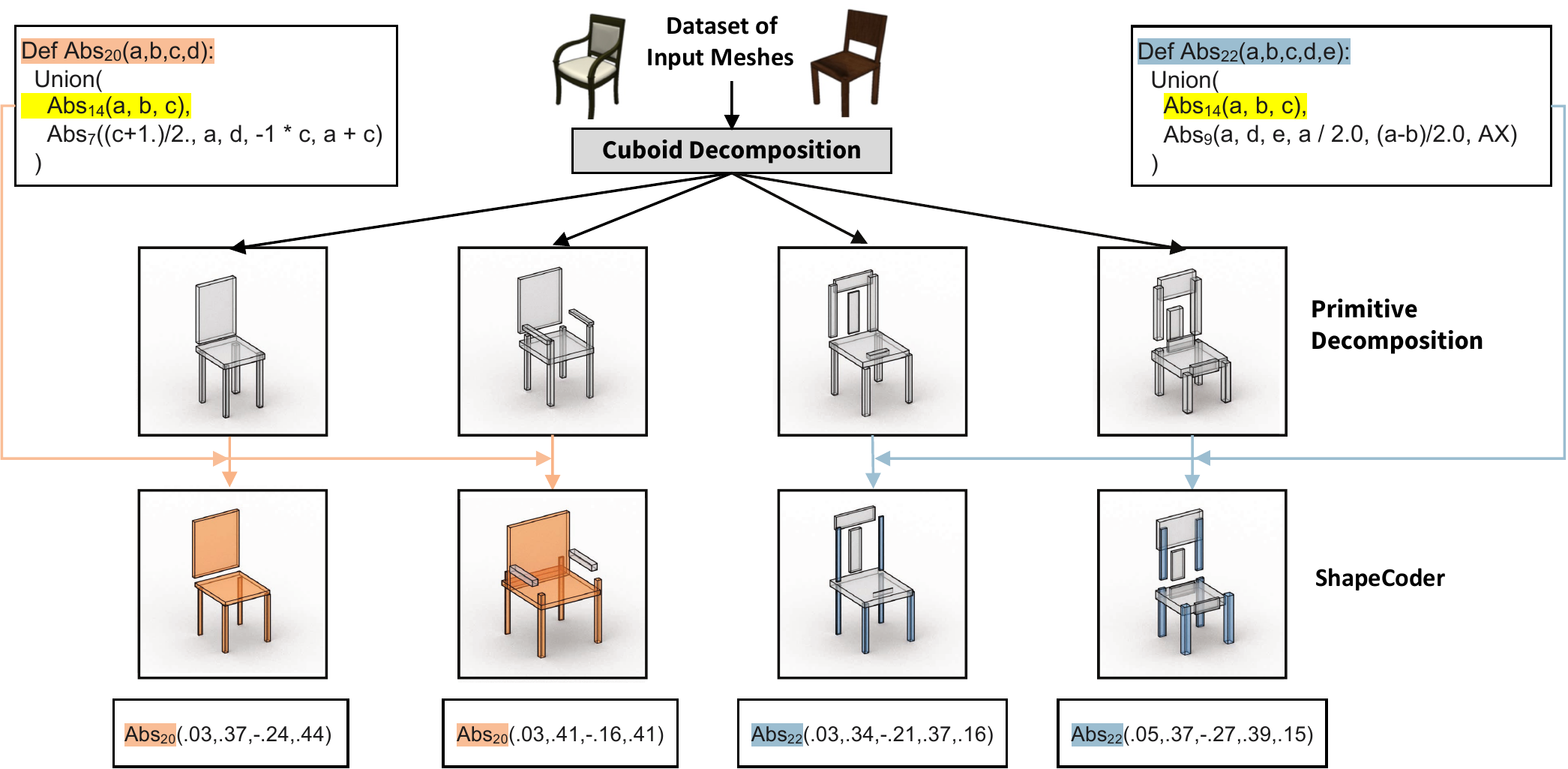}
\caption{We leverage an unsupervised primitive decomposition approach \cite{yang2021unsupcsa} to run \methodname over datasets of 3D meshes. Even on these noisy primitive decompositions, our method still finds high-level, useful abstractions that capture meaningful degrees of shape variation. Interestingly, the two top-level abstractions we show, in orange and blue, both make use of the same abstraction sub-function (highlighted in yellow) to create a four-leg base.
}
\label{fig:unsup_qual}
\end{figure*}

\subsection{ShapeCoder Ablations}
\label{sec:res_ablations}

To evaluate the design decisions behind \methodname, we run an ablation experiment, by tracking how the removal of different components of our method impacts the types of abstractions we discover, and how those abstractions impact the optimization of the objective function. We consider the following ablation conditions:

\begin{itemize}
\denselist
    \item \textit{No Abstraction:} We report the results of running just the wake phase, once, without an abstraction phase.
    \item \textit{Single Iter:} We only run \methodname for a single round.
    \item \textit{No Dream+Wake:} We run multiple rounds of \methodname without access to a recognition network. Instead `naive' programs are used to initialize the algorithm.
    \item \textit{No Semantic Rws:} We remove all of the semantic rewrites associated with our base DSL in the refactor operation. 
    \item \textit{No Conditional Rws:} We replace our conditional rewriting scheme with the `naive' approach described in Section \ref{sec:met_refactor}.
    \item \textit{No Abs Preferences:} We remove the preference weighting \progweight, described in Section \ref{sec:met_obj}.
\denselist
\end{itemize}

We report how these different variants perform in Table ~\ref{table:joint_table}, left, using shapes from the \textit{Chair} category of PartNet.
All ablation conditions lead to worse optimization behavior than our default configuration (bottom row).
Without an abstraction phase, the programs returned from wake can't leverage higher-order functions.
With just a single iteration of \methodname, hierarchical abstractions can't be discovered, and the wake phase can't learn to apply the discovered abstractions more broadly.
When the abstraction phase is run without a dream or wake phase, the method runs into a similar problem, where the abstractions can be underutilized, and won't be integrated into all of the shapes that they could be used to represent.
The semantic rewrites allow e-graphs to represent a large set of equivalent programs that we efficiently search over during refactoring; when we don't consider this large set of equivalent programs, we, once again, under-apply proposed abstractions. 
The importance of our conditional rewrite scheme is made evident by the no conditional rewrite ablation: within the computational budget allotted for this ablation experiment (3 days) the version of \methodname that used the `naive' rewrite scheme failed to finish a complete abstraction phase. 
As such, we report its objective function value at this 3-day cut-off.
Finally, our preference weighting scheme helps \methodname avoid local minima: mostly by down-weighting obviously bad (e.g. too constrained or too general) candidate abstraction functions.

\subsection{Discovering Abstractions from Unstructured Shapes}
\label{sec:res_unstruct}


As an illustrative application of \methodname, we investigate its ability to jointly discover a library of abstraction functions and programs that use those abstractions, when run over a dataset of 3D meshes.
To source this kind of input data, we use a method that performs unsupervised cuboid decomposition of 3D shapes~\cite{yang2021unsupcsa}.
Specifically, we employ this approach to convert sets of ShapeNet meshes into arrangements of unstructured, noisy primitives -- a data format that \methodname can reason over. 
We provide details of this data preprocessing in Appendix \ref{sec:apndx_prim_decomp}

Similar to the experiments in Section \ref{sec:res_discovery}, we construct a dataset of 400 shapes, with primitives produced by this unsupervised algorithm.
We run \methodname over a dataset of chairs sourced from ShapeNet~\cite{shapenet2015}~for three rounds and show results of some of the discovered abstractions in Figure \ref{fig:unsup_qual}. 
Even though the primitive decompositions that \methodname receives are noisy and irregular, it still manages to discover a collection of meaningful abstraction functions that expose higher-order properties and can be applied across instances of the input distribution. 
For instance, the discovered $Abs_{20}$, captures the same fundamental chair structure found by \methodname when run over PartNet annotations ($Abs_{24}$, Figure~\ref{fig:qual_abs}). 
In fact, over the course of 3 rounds, \methodname improves the objective function score by 61\% ($140 \rightarrow 53.9$), which is similar to the quantitative improvement observed when \methodname operates over clean, manually annotated parts.
These results are promising, and indicate that systems like \methodname can be used to discover useful high-level programmatic representations of complex visual phenomena, without reliance on manual annotations.

\subsection{Downstream Benefits of Abstractions}
\label{sec:res_abs_benefits}

\rev{
In this section, we investigate how \methodname's discovered abstractions can benefit downstream applications with two experiments: maintaining validity under perturbations and novel shape synthesis. 
}

\paragraph{Maintaining validity under perturbation} 

\rev{As we aim to discover abstractions that remove extraneous degrees of freedom, we can evaluate success by perturbing degrees of freedom in shape programs, and checking whether they ‘stay in distribution’. 
We take two shape program datasets, where programs are written with or without abstractions, and perturb their parameters under different noise levels. 
Specifically, the noise level modulates the standard deviation of Gaussian noise distributions fit to each parameter slot of each DSL function. 
For each perturbed set of programs, we measure how similar their output executions are to a validation set with Frechet Distance (FD) in the feature space of a pretrained model. 
We report results of this experiment in Table \ref{table:joint_table}, right.
We find that rewriting programs with abstractions discovered by \methodname helps to keep shapes `in distribution' under parameters perturbations, which is an important property for goal-directed editing tasks.}

\paragraph{Novel Shape Synthesis}
\rev{
We evaluate if generative models that learn to write novel shape-programs benefit from training over programs that have been rewritten with discovered abstractions.
For this experiment, we use the PHI procedure (Section \ref{sec:res_qual}) to construct a dataset of 3600 chair-programs written with \methodname  discovered abstractions. 
We use this dataset to train an auto-regressive network, a Transformer decoder, that learns to generate sub-programs conditioned on a canvas that tracks the execution output of previously predicted program parts (Appendix \ref{sec:apndx_gen_model}) 
To synthesize novel shapes, the network starts with a blank canvas, and then gradually builds up a complex program by iteratively sampling expressions, and adding their outputs to the canvas, until a \texttt{STOP} token is predicted.}

\begin{figure}[t!]
 \includegraphics[width=.8\linewidth]{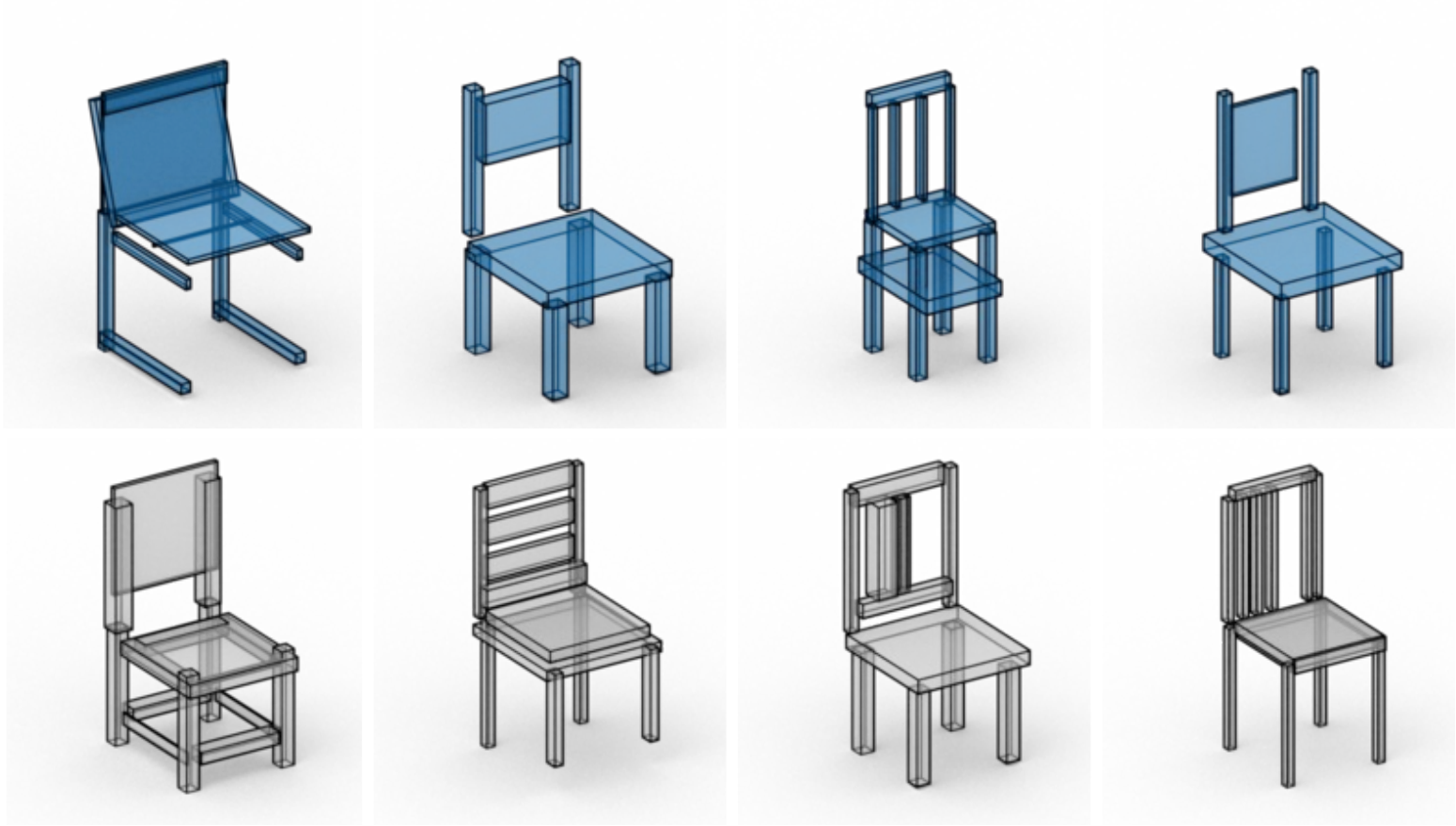}
    \caption{ Sampled programs (top) from a generative model that writes programs containing abstractions, along with nearest neighbors (bottom).}
    \label{fig:gen_qual}
\end{figure}

\rev{We visualize outputs of this model in Figure~\ref{fig:gen_qual}.
Qualitatively, we find that this model can create new shapes not observed from the training set, that clearly stay within the training-distribution. 
Quantitatively, we compare the outputs of this model against an ablated version that trains over programs without abstractions, and find that learning over programs written with abstractions improves Frechet Distance (against a validation set) from 17.1 to 13.8, a 19\% improvement.
Moreover, generative models of visual programs that learn over abstractions are particularly attractive, because the programs they output have less extraneous degrees of freedom, and will be better suited for downstream tasks. }

\section{Conclusion}

We have presented \methodname, a system capable of discovering visual program abstractions in a collection of shapes represented as unstructured primitives. Our method does not require any additional supervision such as ground truth programs, any specific ordering of program operations, or any program curriculum. 
We have shown that \methodname discovers high-level abstractions, that result in significant compression, on domains that other state-of-the-art methods cannot handle. 
\methodname can find programs that use these abstractions to explain shapes not observed during optimization, compactly.
Finally, we demonstrated the flexibility of \methodname by showing that it can discover useful abstractions, that capture meaningful degrees of freedom when run over noisy primitive decompositions produced by an unsupervised method.

\subsection{Limitations and Future Work} While \methodname is the first method to discover non-trivial program abstractions directly from unstructured primitives, it does have some important limitations:

\paragraph{(i) Redundant abstractions.}
We find multiple abstractions that explain the same concept. While these can be seen as structural variations for the same semantic concept (e.g. pedestal chair bases and four-leg chair bases), the abstracted programs can feel redundant for downstream tasks. This is hard to avoid as, at present, we do not `execute' the programs to compare their geometric output. In the future, we want to explore `conditional' or `probabilistic' abstractions. For instance, a chair base abstraction could expand into either a pedestal base or a four-leg base, depending on either a discrete input parameter, or a given probability for each variation.

\paragraph{(ii) Unsaturated e-graphs.}

\rev{
For complicated input expressions, it can be computationally infeasible to fully saturate e-graphs, as they lack the ability to efficiently represent associativity-commutativity constraints. 
While \methodname doesn't offer a direct solution to this issue, our use of conditional rewrites avoids inserting extraneous parametric operation nodes.
This helps to alleviate exponential blowup, and allows \methodname to explore a much richer range of possible program structures than prior work.
Despite this, we cannot always saturate our e-graphs within the allotted computational budget.
}
This implies that some possibly useful rewrites go unexplored and never get appended to the abstraction library. 
One possibility is to amortize the integration stage with neural components: either by learning rewrites (e.g., using a reward structure in a reinforced learning setup), by learning which parts of the e-graph to expand, or by putting the burden of `large' rewrites on a learned module, rather than the e-graph. However, training such modules in an unsupervised setup requires further research.

\paragraph{(iii) Bottom-up wake network.}
\methodname's recognition network (used in the wake phase) solves sub-problems that are stitched together through combinator operations.
A downside of this design decision is that the recognition network must be retrained whenever the library version changes.
Further, as the network does not predict an entire program in one-shot, inference can be expensive to run, and there is less consistency in how programs will be inferred across a dataset.
Replacing this bottom-up network with a top-down network would be more challenging. 
Still, it would open up other possibilities, such as removing the need for the input data to be represented as collections of primitives.

\vspace{1em}
\rev{Looking forward, we believe that \methodname should be helpful for many other visual programming domains, beyond the 2D and 3D shape grammars we consider in this report.  
\methodname requires the following domain attributes: \textit{(a)} the language is functional, \textit{(b)} it contains a combinator operation (e.g. \texttt{Union}), and  \textit{(c)} visual inputs can be decomposed into primitive types. 
In fact, properties \textit{(b)} and \textit{(c)} are only needed for the wake phase, so this requirement could be relaxed by using program inference networks that consume `raw' visual data. 
Sketches, CSG, SVG, and even shader programs could make good matches for future explorations.}
For the first time, \methodname provides the ability to perform program abstraction discovery directly on unstructured collections of primitives, reducing the burden of collecting, annotating, and grouping shape categories.

\begin{acks}

We would like to thank Srinath Sridhar and the anonymous reviewers for their helpful suggestions. 
Renderings of shape programs were produced using the Blender Cycles renderer. 
This work was funded in parts by NSF award \#1941808, a Brown University Presidential Fellowship, and an ERC grant (SmartGeometry).
Daniel Ritchie is an advisor to Geopipe and owns equity in the company. 
Geopipe is a start-up that is developing 3D technology to build immersive virtual copies of the real world with applications in various fields, including games and architecture.
\end{acks}

\bibliographystyle{ACM-Reference-Format}
\bibliography{main, EG2013STAR}
\appendix
\section{Shape Grammar}
\label{sec:apndx_grammar}

\paragraph{3D Shape Grammar}
\label{sec:apndx_3d_lang}
Below we detail our 3D shape grammar:\\

\begin{tabular}{|l|}
\hline
\textit{START}  $\xrightarrow{}$  \textit{SHAPE} \\
\textit{SHAPE} $\xrightarrow{}$ \texttt{Union}(\textit{SHAPE}, \textit{SHAPE}) | \\
\hspace{3.5em}      \texttt{SymRef}(\textit{SHAPE}, AXIS) | \\
\hspace{3.5em}      \texttt{SymTrans}(\textit{SHAPE}, \textit{AXIS}, \textit{INT}, \textit{FLOAT}) | \\
\hspace{3.5em}      \texttt{Rotate}(\textit{SHAPE}, \textit{AXIS}, \textit{FLOAT}) | \\
\hspace{3.5em}      \texttt{Move}(\textit{SHAPE}, \textit{FLOAT}, \textit{FLOAT}, \textit{FLOAT}) | \\
\hspace{3.5em}      \texttt{Cuboid}(\textit{FLOAT}, \textit{FLOAT}, \textit{FLOAT}); \\
\textit{AXIS} $\xrightarrow{}$ \texttt{AX} | \texttt{AY} | \texttt{AZ}; \\
\textit{INT} $\xrightarrow{}$ [1, 6]; \\
\textit{FLOAT} $\xrightarrow{}$ $\mathrm{Prim}_{ij}$ | -1 | 0 | 1 | 2 | \\
\hspace{3.5em} \texttt{Add}(\textit{FLOAT}, \textit{FLOAT}) | \texttt{Sub}(\textit{FLOAT}, \textit{FLOAT}) \\
\hspace{3.5em} \texttt{Mul}(\textit{FLOAT}, \textit{FLOAT}) | \texttt{Div}(\textit{FLOAT}, \textit{FLOAT}); \\
\hline
\end{tabular}
\newline

We italicize all non-terminal parts of the grammar, and explain what the terminal operators in the language do (non-italicized). 
\texttt{Union} combines two sub-shapes together. 
\texttt{SymRef} is a symmetry reflection across an axis. 
\texttt{SymTrans} is a symmetry translation over an axis, that creates a specified number of copies, up to a specified distance. 
\texttt{Rotate} specifies an Euler angle rotation about an axis.
\texttt{Move} moves a cuboid by a specified amount.
\texttt{Cuboid} instantiates a cuboid with the specified dimensions. 
Axes can be either the X, Y, or Z axis.
Ints can be an integer between 1 and 6.
Floats can be either be sourced from a primitive parameter of an input scene ($\mathrm{Prim}_{ij}$), be a constant, or the result of a parametric operation. 

\paragraph{2D Shape Grammar}
\label{sec:apndx_2d_lang}

Below we detail our 2D shape grammar:\\

\begin{tabular}{|l|}
\hline
\textit{START}  $\xrightarrow{}$  \textit{SHAPE} \\
\textit{SHAPE} $\xrightarrow{}$ \texttt{Union}(\textit{SHAPE}, \textit{SHAPE}) | \\
\hspace{3.5em}      \texttt{SymRef}(\textit{SHAPE}, AXIS) | \\
\hspace{3.5em}      \texttt{SymTrans}(\textit{SHAPE}, \textit{AXIS}, \textit{INT}, \textit{FLOAT}) | \\
\hspace{3.5em}      \texttt{Move}(\textit{SHAPE}, \textit{FLOAT}, \textit{FLOAT}) | \\
\hspace{3.5em}      \texttt{Rect}(\textit{FLOAT}, \textit{FLOAT}); \\
\textit{AXIS} $\xrightarrow{}$ \texttt{AX} | \texttt{AY} ; \\
\textit{INT} $\xrightarrow{}$ [1, 4]; \\
\textit{FLOAT} $\xrightarrow{}$ $\mathrm{Prim}_{ij}$ | -1 | 0 | 1 | 2 | \\
\hspace{3.5em} \texttt{Add}(\textit{FLOAT}, \textit{FLOAT}) | \texttt{Sub}(\textit{FLOAT}, \textit{FLOAT}) \\
\hspace{3.5em} \texttt{Mul}(\textit{FLOAT}, \textit{FLOAT}) | \texttt{Div}(\textit{FLOAT}, \textit{FLOAT}); \\
\hline
\end{tabular}
\newline

This is a simplified version of our 3D grammar, where the rotation command has been removed, and all 3D parameterizations are replaced with 2D parameterizations.

\section{Implementation Details}
\label{sec:apndx_impl_dets}

We provide implementation details for \methodname below.
For all experiments in Section \ref{sec:results} we set \absnum = 20 and \dreamnum = 10000. 

\subsection{Objective Function Weights}
\label{sec:apndx_obj_fn}

We use the following weights for \weight in \methodname's objective function (Section \ref{sec:met_obj}): float tokens are 2.0, shape-returning function tokens are 1.0, float-returning function tokens are 0.1 (i.e. parametric operations), and categorical tokens (including integers) are 0.5 .
Additionally we set the geometric error weight, $\lambda_e$, to be 10.

For the function weighting scheme \progweight, described in Section \ref{sec:met_obj} and ablated in Section \ref{sec:res_ablations}, \methodname employs the following logic.
The base cost of adding a new abstraction \function into \library is 0.25, but this value can be modulated within the range of 0.125 to 0.5 based on properties of \function.
The presence of parametric expressions in \function decrease~\progweight.
Too many input parameters in \function increases \progweight, where more than~6 parameters starts to incur penalties, and abstractions with more than 10 input parameters are rejected outright.
We decrease \progweight for doubleton abstractions (those that use multiple sub-functions), and increase \progweight for singleton abstractions that use a single sub-function.
Finally, if \function is found to be used very infrequently over \programs, less than~1\% observation rate, then we also reject \function outright.

\subsection{Geometric Error Function}
\label{sec:apndx_geom_err}

The objective function (Section \ref{sec:met_obj}) uses a geometric error function~$err$ that compares how closely an executed expression \expr from \library matches a target shape \datapoint. 
As this error function is used extensively in the wake phase (Section \ref{sec:met_wake}), it checks for partial solutions.
Say executing \expr creates a set of primitives $prim_\expr$, and \datapoint contains primitives~$prim_\datapoint$.
First our geometric error functions finds an optimal mapping from primitives in $prim_\expr$ to some primitive in $prim_\datapoint$. 
Mechanically, we construct a distance matrix of size $|prim_\expr|\times~ |prim_\datapoint|$, that calculates a domain-specific distance metric between each pair of input and target primitives (explained later). 
For any pair of primitives whose distance is above a user-defined maximum error threshold, we set their paired distance to an arbitrarily high value (10000).
We use the Hungarian matching algorithm to find an optimal match over this distance matrix.
If none of the paired matches between $prim_\expr$ and $prim_\datapoint$ have distance over 10000, then the match is valid, and the total error incurred by \expr for \datapoint is simply the sum of all entries in the distance matrix involved in this optimal match.

During the integration phase (Section \ref{sec:met_integration}), we can modify this approach to check for a \textit{program} \program that explains \datapoint, by enforcing that the distance matrix must be square. Whenever this condition is not met, it means that there is a mismatch in the number of primitives created by \program, and the number of primitives expected in the target shape \datapoint, so \program is invalid. 

\paragraph{2D geometric distance} 
Each primitive (rectangle) is represented as 4 parameters: width, height, x position, and y position.
To find the distance between two primitives, we take the average of the absolute differences between each parameter slot. 
The maximum allowable error threshold is set to 0.05.

\paragraph{3D geometric distance}
Each primitive (cuboid) is represented as 9 parameters: dimensions, position, Euler angle rotations. 
To find the distance between two primitives, we calculate the corner positions of each cuboid, and record the Hausdorff distance between the two sets of points.
The maximum allowable error threshold is set to 0.1 .

\subsection{Recognition Network}
\label{sec:apndx_rec_net}

Our recognition network uses a Transformer decoder backbone architecture with causal masking.
We allow it to condition on up to 16 primitives (where each primitive will contribute $K$ tokens), and fix its max prediction length to be 32.
It uses 2 attention blocks, with 8 heads in each block, and a hidden dimension of 128.
Training uses a batch size of 64, dropout of 0.5, and a learning rate of .0001. 
Each dream phase (Section \ref{sec:met_dream}) trains the recognition network for a maximum of 300 epochs, where early stopping is performed on a validation set of held-out dreams (10\% of samples).

\subsection{Dream Creation}
\label{sec:apndx_dream}

\paragraph{Sampling library functions}
During the dream phase (Section \ref{sec:met_dream}), \methodname randomly samples instantiations of library functions to train the recognition network. Some dreams are visualized in Figure~\ref{fig:qual_abs}.
For each discrete decision needed to parameterize a function~\function, we find all tokens in \library that type-match, and uniformly sample from this distribution.
Float-typed tokens are represented as mixtures of Gaussians distributions (max 3 mixture components). 
These distributions are designed to broadly reflect reasonable values for certain parameter slots in the base DSL. 
For instance, the first float parameter slot in the `Move' operator is associated with x-axis positioning, so we design a trimodal mixture distribution with the following properties:
it has a  0-centered dominant component, and then two minor components placed to the left and right of the origin.
These distributions don't meaningfully change the performance of the recognition model, as it gets to trains on a massive amount of samples, but it does speed up the rate at which we can find valid dreams under our rejection criteria (explained below).
When sampling dreams for abstraction functions, the parameter inputs in the abstraction inherent the distributions of their child sub-functions.

\paragraph{Dream rejection criteria} 
We use simple checks to validate that randomly sampled dreams produce meaningful training data, and reject any dreams that don't meet the following criteria.
All primitives must have positive dimensions.
The corners of all primitives must be within the allotted scene bounding volume $[-1,1]^n$, with a~10\% leniency threshold.
At least 50\% of each primitives area must be visible (i.e. not contained within another primitive). 
Each primitive must be bigger than a specified threshold: 0.005 area of 2D, .00025 volume for 3D.
Dreams cannot contain more than 16 primitives. 
Dreams cannot use redundant operations, for instance, applying two \texttt{Move} commands in a row.

\paragraph{Forming composite scenes}
\methodname's recognition network trains on composite scenes, that are formed by sampling function-specific dreams and combining them together.
To form a composite scene, we sample a random integer $k$ from [1,4], sample $k$ functions from the set of all library functions that have not been represent in \dreamnum dreams, and choose a random dream from each chosen function.
Additionally, with 50\% chance, we add distractor primitives into the composite scene.
Distractor primitives are sourced by randomly sub-sampling primitives found in some \datapoint$\in$~\dataset.
To encourage the recognition network to be position invariant, we optionally sample a \texttt{Move} operation (with 50\% frequency) and apply it over the primitives created by a function-specific dream.
Note that this \texttt{Move} operation is not included in the target expression, so the recognition network must become invariant to where the target primitives show up in the composite scene.

\subsection{Combining Wake Programs}
\label{sec:apndx_comb_wake}

As discussed in Section \ref{sec:met_wake}, programs discovered in round $r$'s wake phase need to be combined with programs discovered in rounds before $r$. 
Here we detail how \textit{combine} is implemented.

Assume we are in the wake phase of round $r$, $r > 0$.
For some~\datapoint~$\in$~\dataset there is currently some program entry in \programs, $\program_c$.
Using a \textit{split} function, that recursively removes combinator operations from a program, we can convert $\program_c$ into a set of expressions in \library:  

\noindent
$split(\program_c)~=~E_c~=~\{e_c^0,..., e_c^{|E_c|}\}$.
When executed, each $e_c^i$ will create a set of primitives, $prim_{c}^{i}$, that is a subset of the primitives in \datapoint.
\methodname keeps track of all such previous expressions associated with~\datapoint in a data-structure $Q_d$, sourced from either the wake or integration phases. 

The wake inference procedure uses the recognition network to prediction a new program in round $r$, $\program_r$, for \datapoint. 
We decide what program \program should be kept in \programs by constructing 4 program variants, and keeping the one that minimizes \objective.
The variants we consider are as follows.
\textit{(i)} Use $\program_c$.
\textit{(ii)} Use $\program_r$ (note this variant will always be chosen if $r=0$).
\textit{(iii)} Greedily merge $\program_r$ into $\program_c$. 
To do this, we first compute $split(\program_r) = E_r = \{e_r^0,..., e_r^{|E_c|}\}$.
Then for each~$e_r^i$, we find~$prim_{r}^{i}$, and see if there is a set of matching instances in~$E_c$,~$M$, such that $prim_{r}^{i} = \{ prim_{c}^{j}~ \mathrm{for}~j \in M \}$.
If $M$ exists, then we compare the cost under \objective of $e_r^i$ versus the sum of each $e_{c}^{j}$ (with $|M|-1$ combinator calls): if $e_r^i$ improves \objective then each $e_{c}^{j}$ is removed from~$E_c$, and $e_r^i$ is added into $E_c$.
\textit{(iv)} Greedily construct an entirely new program from $Q_d$. 
First $E_r$ is added into $Q_d$. 
Then $Q_d$ greedily creates a new program by initializing $E_n$ (to be empty) and repeating the following steps: find the \textit{cost} of each \expr in $Q_d$, take the minimum cost expression~$e^*$ and add it into $E_n$, and temporarily remove all other entries of $Q_d$ that have nonzero overlap with $prim_e^*$. 
This is repeated until $E_n$ contains expressions that cover all primitives in \datapoint.

After these four program variants have been created (where in \textit{(iii)} and \textit{(iv)} combinator operations are applied over $E_c$ and $E_n$ respectively), the variant with the minimum score under \objective is kept in \programs. 
Finally, we note that some extra logic is required to ensure that~$Q_d$ and $\program_c$ are kept up-to-date.
Whenever the integration phase tries removing a function \function from \library, 
all expressions in $Q_d$ that use~\function are temporarily removed.
Moreover if \function appears in $\program_c$, then the greedy search in \textit{(iv)} is used to find replacement expressions for $\program_c$.

\subsection{Preference Ordering of Parametric Relationships}
\label{sec:apndx_ord_op}

The proposal phase (Section \ref{sec:met_proposal}) generates candidate abstractions using a greedy search.
These candidate abstractions contain parametric expressions. 
Below we detail the preference ordering we use to search for matching parametric expressions with respect to a sampled cluster.

The choice of which parametric expression to propose is always made in the context of a cluster, that contains a structure and a group of parameterizations.
As we are filling in slots for the candidate abstraction, we may have already instantiated free variables that were used in previous slots.
To find a possible expression for the current parameter slot, we reason over the free variables previously instantiated.
We iterate through a preference ordering that considers increasingly complex parametric expressions over previous variables: expressions with only constants, then one variable expressions, two variable expressions, and finally three variable expressions. 
The set of all expressions under \library that contain $n$ variables can be found by calculating the cross-product of (i) all parametric operator combinations that would require $n$ variables with (ii) all ordered sequences of $n$ previously instantiated variables.
To avoid overfitting, we limit the possible constants we consider (just 0 for our shape grammars). 
For each expression, we check which members of the cluster are covered by that expression. 
Once we find a set of expressions that collectively cover all instances within the cluster, we break out of this loop early.
This procedure creates a large set of possible expressions (visualized in Figure \ref{fig:proposal}), from which one is chosen according to the \textit{score} function.

\subsection{E-graphs}
\label{sec:apndx_egraphs}

Our refactor operation (Section \ref{sec:met_refactor}), implements e-graphs using the Egg library \cite{2021-egg}.
Egg provides support for defining a DSL, rewrite operations, and a cost function, that can be used by an extraction operation. 
Egg provides an interface for defining rewrites that reason over conditional logic, but they cannot be directly applied for our use case. 
Our version of conditional rewrites requires that each rewrite has access to a shared e-class-to-real-value mapping, so we build out this feature.
Maintaining this mapping requires dummy rewrite operations, that check for structural matches for various parametric operations, and update the mapping, without changing the structure of the e-graph.
When we first instantiate an e-graph, we apply dummy rewrites that match on each float variable, $V_i$, and adds an entry for $V_i$ into the mapping. 
Then, during each rewrite round, after applying all semantic and abstraction rewrites, we apply all dummy rewrites, to ensure the mapping is up-to-date (this handles the blue \texttt{Mul} e-class from Figure \ref{fig:spad_egraph}).
For each domain, we provide Egg with a set of semantic rewrites that express domain-specific semantic preserving transformations. 
There are 25 such rewrites for 3D, and~16 such rewrites for 2D. We ablate the importance of including these semantic rewrites in our ablation experiment (Section \ref{sec:res_ablations}). 

\subsection{Unsupervised Primitive Decomposition}
\label{sec:apndx_prim_decomp}

As described in Section \ref{sec:res_unstruct}, we make use of an unsupervised cuboid decomposition method, so that we can apply \methodname to shapes from datasets that contain only meshes. 
We use the approach described by \cite{yang2021unsupcsa}, using their released pretrained models to predict cuboid decompositions over chairs from their test set.
We compile a dataset of 400 such predictions, and parse these output predictions into a primitive representation compatible with our method. 
This conversion procedure performs a few minor filtering steps, rejecting scenes that contain more than 12 cuboids (we found these often were noisy predictions) and snapping cuboids to be axis-aligned whenever their Euler angles were within a 0.05 threshold of 0 or $2\pi$.

\subsection{Generative Model for Programs}
\label{sec:apndx_gen_model}

We provide details for the generative model described in Section~\ref{sec:res_abs_benefits}. 
This model is capable of synthesizing novel 3D shapes.
We implement our generative model as a Transformer decoder, with causal masking. 
It uses a CNN to encode a shape voxelization into an embedding vector, which conditions the Transformer that autoregressively predicts tokens from \library. 
The network starts with a blank scene, iteratively predicts an expression \expr from \library, and adds it back into the scene (which will be encoded by the CNN in the next time-step). 
This process is repeated until a special `STOP' token is predicted.

We source training data for this model by running our post hoc inference procedure (Section \ref{sec:res_qual}) over a dataset of 3600 chairs, to form a program dataset \programs. 
For each epoch, we randomize expression ordering by applying \textit{split} (Section \ref{sec:apndx_comb_wake}) to each \program$\in$~\programs, shuffling the expressions found by \textit{split}, and treating every (previous expressions, next expression) tuple as an independent training example.
We use teacher-forcing and maximum likelihood updates to train the generative model. 
We train the model for 4000 epochs. It has 8 Transformer layers, 16 heads, a hidden size of 256. We train with a batch size of 64, dropout of 0.1, and a learning rate of 0.0005 .
At inference time, we use nucleus sampling (top 90\%) to predict expressions from the networks probabilities. 
The `without abstractions' version we compare against has exactly the same setup, except the post-hoc inference procedure was run using the starting \library version (not the one discovered by \methodname).

\end{document}